\documentclass[final,nomarks]{dmtcs-episciences}

\usepackage[utf8]{inputenc}
\usepackage{subfigure}
\publicationdetails{22}{2021}{2}{5}{6002}

\usepackage{amsmath}
\usepackage{amsthm,amsfonts,amssymb}
\newcommand{\comm}[1]{#1}

\newtheorem{Le}{Lemma}
\newtheorem{Co}{Corollary}

\newtheorem{Pro}{Proposition}
\theoremstyle{definition}

\usepackage[english]{babel}
\usepackage[T1]{fontenc}
\usepackage{times}
\usepackage[utf8]{inputenc}
\usepackage{vmargin}
\usepackage[table]{xcolor}
\usepackage{rotating}
\usepackage[utf8]{inputenc}
\usepackage[normalem]{ulem}
\usepackage{enumitem}
\usepackage{tikz}
\usetikzlibrary{patterns}

 \def\A{0.5cm}
 
 \def\C{2.5cm}
 
 \def\E{4.5cm}
 
 \def\G{6.5cm}
 
 \def\I{8.5cm}
 
 \def\K{10.5cm}
 \def\L{11.5cm}
 \def\M{12.5cm}
 
 \def\O{14.5cm}\def\Q{16.5cm}\def\S{18.5cm}
\def\U{20.5cm}\def\V{21.5cm}\def\W{22.5cm}\def\Y{24.5cm}\def\ZZ{26.5cm}
\def\ZZ{26.5cm}\def\Za{28.5cm}\def\Zb{30.5cm}\def\Zc{32.5cm}

\def\styleGrille{densely dotted}
\author{
Jean-Luc Baril\affiliationmark{1}
  \and Carine Khalil \affiliationmark{1}
  \and Vincent Vajnovszki\affiliationmark{1}} 

\title{Catalan words avoiding  pairs of length three patterns}

\affiliation{
  LIB, Universit\'e de Bourgogne Franche-Comt\'e, Dijon, France
}

\received{2019-12-25}

\revised{2021-03-31} 

\accepted{2021-04-01}

\begin{document}
\publicationdetails{22}{2021}{2}{5}{6002}

\maketitle
\begin{abstract}
Catalan words are particular growth-restricted words counted by the eponymous integer sequence. 
In this article we consider Catalan words avoiding a pair of patterns of length $3$, 
pursuing the recent initiating work of the first and last authors and of S. Kirgizov
where (among other things) the enumeration of Catalan words avoiding a patterns
of length $3$ is completed.
More precisely, we explore systematically the structural properties of the sets of words under consideration and give enumerating results by 
constructive bijections or bivariate generating functions with respect to the length and descent number. 
Some of the obtained enumerating sequences are known, and thus the corresponding results establish new combinatorial interpretations for them.
\end{abstract}

\section{Introduction and notation}

\medskip

Catalan words are particular growth-restricted words 
and they represent still another combinatorial class counted by the Catalan numbers, 
see for instance \cite[exercise 6.19.u, p. 222]{Stanley}.
This paper contributes to a recent line of research on classical pattern avoidance
on words subject to some growth restrictions (for instance, ascent sequences \cite{Baxter_Pudwell_15,Duncan}, inversion sequences
\cite{CMSW,Mansour_Shatuck,Yan_Lin}, 
restricted growth functions \cite{CDDGGPS,Lin_Fu}) by investigating connections between sequences on the On-line Encyclopedia of Integer Sequences \cite{OEIS}  and Catalan words avoiding two patterns of 
length $3$. 

Through this paper we consider words over the set of non-negative integers
and we denote such words by sequences (for instance $w_1w_2\dots w_n$) or by italicized boldface letter
(for instance $\boldsymbol{w}$ and $\boldsymbol{u}$).
The word $\boldsymbol{w}=w_1w_2\ldots w_n$ is called a
{\it Catalan word} if 
$$w_1=0\mbox{ and } 0\leq w_i\leq w_{i-1}+1\mbox{ for } i=2,3,\ldots ,n.$$
Catalan words are in bijection with maybe the most celebrated combinatorial class having the same
enumerating sequence: Dyck paths\footnote{A Dyck path is a path in the first quadrant 
of the plane
which begins at the origin, ends at $(2n, 0)$, and consists of up steps $(1, 1)$ and down steps
$(1,-1)$.}.
Indeed, in a length $2n$ Dyck path collecting for the up steps the 
ordinates of their starting points we obtain a length $n$ Catalan word, and this construction is a bijection.
See Figure~1 where this bijection is depicted for an example.
We denote by $\mathcal{C}_n$ the set of length $n$ Catalan words and $c_n=|\mathcal{C}_n|$ is the $n$th Catalan number $\frac{1}{n+1} {2n\choose n}$.

\medskip
A {\it pattern} is a word with the property that if $i$ occurs in it, then so does $j$, for any $j$ with $0\leq j<i$.
A pattern $\pi=\pi_1\pi_2\dots \pi_k$ is said to be contained in the word  $\boldsymbol{w}=w_1w_2\ldots w_n$, $k \leq n$, if there is a  sub-word of $\boldsymbol{w}$, $w_{i_1}w_{i_2}\dots w_{i_k}$, order-isomorphic with $\pi_1\pi_2\dots \pi_k$. 
If $\boldsymbol{w}$ does not contain $\pi$, we say that $\boldsymbol{w}$ {\it avoids} $\pi$,
see for instance Kitaev's seminal book \cite{Kitaev2011} on this topic.

For a pattern $\pi$, we denote by $\mathcal{C}_n(\pi)$ the set of length $n$ Catalan words avoiding $\pi$, 
and  $c_n(\pi)=|\mathcal{C}_n(\pi)|$ is the cardinality of $\mathcal{C}_n(\pi)$ and $\mathcal{C}(\pi)=\cup_{n\geq0}\mathcal{C}_n(\pi)$.
For example, $\mathcal{C}_n(101)$ is the set of length $n$ Catalan words avoiding $101$, that is,
the set of words $\boldsymbol{w}$ in $\mathcal{C}_n$ such that there are no $i$, $j$ and $k$,
$1\leq i<j<k\leq n$, with $w_i=w_k>w_j$.
So, $\mathcal{C}_4(101)=\{ 
0 0 0 0,
 0 0 0 1, 
 0 0 1 0, 
 0 0 1 1, 
 0 0 1 2, 
 0 1 0 0,$
$ 0 1 1 0, 
 0 1 1 1, 
 0 1 1 2, 
 0 1 2 0,
 0 1 2 1, $
 $
 0 1 2 2, 
 0 1 2 3
\}$.
Likewise, if $\pi$ is the set of patterns $\{\alpha,\beta,\dots\}$, then $\mathcal{C}_n(\pi)$
and $\mathcal{C}_n(\alpha,\beta,\dots)$ denote both
the set of length $n$ Catalan words avoiding each pattern in $\pi$; and 
$c_n(\pi)=c_n(\alpha,\beta,\dots)$ and 
$\mathcal{C}(\pi)=\mathcal{C}(\alpha,\beta,\dots)$ have similar meaning as above.
A {\it descent} in a word $\boldsymbol{w}=w_1w_2\dots w_n$ is a position $i$, $1\leq i\leq n-1$,
with $w_i>w_{i+1}$.
The (ordinary) generating function of a set of pattern avoiding Catalan words $\mathcal{C}(\pi)$
is the formal power series
$$C_\pi(x)=\sum_{n\geq 0}c_n(\pi)x^n=\sum_{\boldsymbol{w}\in\mathcal{C}(\pi)}x^{|\boldsymbol{w}|},
$$
where $|\boldsymbol{w}|$ is the length of the word $\boldsymbol{w}$.
In our case of generating function approach for counting classes of pattern avoiding Catalan words
we consider the descent number as an additional statistic obtaining `for free' the bivariate generating function
$$C_\pi(x,y)=\sum_{\boldsymbol{w}\in\mathcal{C}(\pi)}x^{|\boldsymbol{w}|}
y^{\mathrm{des}(\boldsymbol{w})},
$$
where $\mathrm{des}(\boldsymbol{w})$ is the number of descents of $\boldsymbol{w}$. With these notations,
the coefficient of $x^ny^k$ in $C_\pi(x,y)$ is the number of Catalan words of length $n$ avoiding $\pi$ 
and having $k$ descents, and for a set $\mathcal{S}$ of Catalan words $S(x)$ and $S(x,y)$ have similar meaning. 

For a word $\boldsymbol{w}=w_1w_2\dots w_n$ and an integer $a$, we denote by 
$(\boldsymbol{w}+a)$ the word
obtained from $\boldsymbol{w}$ by increasing by $a$ each of its entries, that is, the word
$(w_1+a)(w_2+a)\cdots (w_n+a)$.
In our constructions we will often make use of two particular families of 
Catalan words: those avoiding $10$ (i.e., with no descents) 
and we call these words {\it weakly increasing} (or {\it w.i} for short) {\it Catalan words};
and those  avoiding $00$ (and thus necessarily avoiding $10$)
and we call these words {\it strictly increasing} (or {\it s.i} for short) {\it Catalan words}.
It is easy to see that for each length $n\geq 1$ there are $2^{n-1}$ w.i. Catalan words 
and one s.i. Catalan word.

The remaining of this paper is structured as follows. In the next section
we characterize pattern avoiding ascent sequences which are Catalan words, establishing ties with 
some similar enumerative results for ascent sequences in \cite{Baxter_Pudwell_15}.
In Section \ref{Sect_2_3} we consider classes of Catalan words avoiding both a length two and a length three pattern. In the next sections 
we discuss  Catalan words avoiding two patterns of length three, in increasing order of their complexity:
obvious cases (Section \ref{Sect_trivial}), cases counted via recurrences (Section \ref{Sect_recure})  and cases counted via generating functions (Section \ref{Sect_gen});
these results are summarized in Table \ref{table2}.
We conclude with some remarks and further research directions.

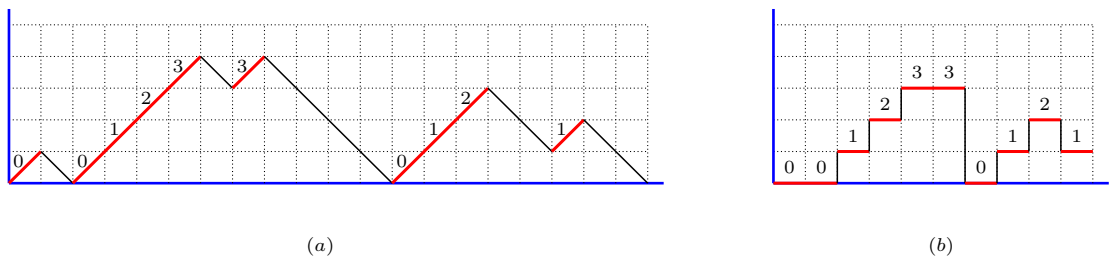
\begin{figure}[h]

        \begin{center}
\begin{tikzpicture}[scale=0.21]
            \draw[\styleGrille] (\A,\A)-- (40.5cm,\A);
             \draw[\styleGrille] (\A,\E)-- (40.5cm,\E);
              \draw[solid,line width=0.35mm,color=blue](\A,\A)-- (41.5cm,\A);
               \draw[\styleGrille] (\A,\G)-- (40.5cm,\G);
               \draw[\styleGrille] (\A,\I)-- (40.5cm,\I);
              \draw[\styleGrille] (\A,\K)-- (40.5cm,\K);
              \draw[\styleGrille] (\A,\C)-- (40.5cm,\C);
            \draw[solid,line width=0.35mm,color=blue] (\A,\A) -- (\A,\L);
             \draw[\styleGrille] (\C,\A) -- (\C,\K);\draw[\styleGrille] (\E,\A) -- (\E,\K);\draw[\styleGrille] (\G,\A) -- (\G,\K);
             \draw[\styleGrille] (\I,\A) -- (\I,\K);\draw[\styleGrille] (\K,\A) -- (\K,\K);\draw[\styleGrille] (\M,\A) -- (\M,\K);
             \draw[\styleGrille] (\O,\A) -- (\O,\K);\draw[\styleGrille] (\Q,\A) -- (\Q,\K);\draw[\styleGrille] (\S,\A) -- (\S,\K);
             \draw[\styleGrille] (\U,\A) -- (\U,\K);\draw[\styleGrille] (\W,\A) -- (\W,\K);\draw[\styleGrille] (\Y,\A) -- (\Y,\K);

             \draw[\styleGrille] (\ZZ,\A) -- (\ZZ,\K);
             \draw[\styleGrille] (\Za,\A) -- (\Za,\K);
             \draw[\styleGrille] (\Zb,\A) -- (\Zb,\K);
             \draw[\styleGrille] (\Zc,\A) -- (\Zc,\K);
             \draw[\styleGrille] (36.5cm,\A) -- (36.5cm,\K);
             \draw[\styleGrille] (38.5cm,\A) -- (38.5cm,\K);
             \draw[\styleGrille] (40.5cm,\A) -- (40.5cm,\K);
             \draw[\styleGrille] (34.5cm,\A) -- (34.5cm,\K);

            \draw[solid,line width=0.4mm,color=red] (\A,\A)--(\C,\C);
            \draw[solid,line width=0.2mm] (\C,\C)-- (\E,\A);
             \draw[solid,line width=0.4mm,color=red] (\E,\A)-- (\M,\I);
             \draw[solid,line width=0.2mm] (\M,\I)-- (\O,\G);
              \draw[solid,line width=0.4mm,color=red](\O,\G)-- (\Q,\I);
              \draw[solid,line width=0.2mm](\Q,\I) -- (\Y,\A);
              \draw[solid,line width=0.4mm,color=red](\Y,\A) -- (30.5cm,\G);
              \draw[solid,line width=0.2mm](30.5cm,\G) -- (34.5cm,\C);
              \draw[solid,line width=0.4mm,color=red] (34.5cm,\C)-- (36.5cm,\E);
              \draw[solid,line width=0.2mm](36.5cm,\E)--(40.5cm,\A);
              \draw (1.1cm,1.9cm) node {\scriptsize $0$};
              \draw (5.1cm,1.9cm) node {\scriptsize $0$};
               \draw (7.1cm,3.9cm) node {\scriptsize $1$};
               \draw (9.1cm,5.9cm) node {\scriptsize $2$};
               \draw (11.1cm,7.9cm) node {\scriptsize $3$};
               \draw (15.1cm,7.9cm) node {\scriptsize $3$};
                \draw (25.1cm,1.9cm) node {\scriptsize $0$};
                  \draw (27.1cm,3.9cm) node {\scriptsize $1$};
                   \draw (29.1cm,5.9cm) node {\scriptsize $2$};
                     \draw (35.1cm,3.9cm) node {\scriptsize $1$};
                  \draw (20cm,-3.5cm) node {\scriptsize $(a)$};
         \end{tikzpicture}\qquad\qquad
         \begin{tikzpicture}[scale=0.21]
            \draw[\styleGrille] (\A,\A)-- (\U,\A);
             \draw[\styleGrille] (\A,\E)-- (\U,\E);
              \draw[solid,line width=0.35mm,color=blue](\A,\A)-- (\V,\A);
               \draw[\styleGrille] (\A,\G)-- (\U,\G);
               \draw[\styleGrille] (\A,\I)-- (\U,\I);
              \draw[\styleGrille] (\A,\K)-- (\U,\K);
              \draw[\styleGrille] (\A,\C)-- (\U,\C);
            \draw[solid,line width=0.35mm,color=blue] (\A,\A) -- (\A,\L);
             \draw[\styleGrille] (\C,\A) -- (\C,\K);\draw[\styleGrille] (\E,\A) -- (\E,\K);\draw[\styleGrille] (\G,\A) -- (\G,\K);
             \draw[\styleGrille] (\I,\A) -- (\I,\K);\draw[\styleGrille] (\K,\A) -- (\K,\K);\draw[\styleGrille] (\M,\A) -- (\M,\K);
             \draw[\styleGrille] (\O,\A) -- (\O,\K);\draw[\styleGrille] (\Q,\A) -- (\Q,\K);\draw[\styleGrille] (\S,\A) -- (\S,\K);
             \draw[\styleGrille] (\U,\A) -- (\U,\K);
            \draw[solid,line width=0.4mm,color=red] (\A,\A)--(\E,\A);
            \draw[solid,line width=0.2mm] (\E,\A)-- (\E,\C);
             \draw[solid,line width=0.4mm,color=red] (\E,\C)-- (\G,\C);
             \draw[solid,line width=0.2mm] (\G,\C)-- (\G,\E);
              \draw[solid,line width=0.4mm,color=red](\G,\E)-- (\I,\E);
              \draw[solid,line width=0.2mm](\I,\E) -- (\I,\G);
              \draw[solid,line width=0.4mm,color=red](\I,\G) -- (\M,\G);
              \draw[solid,line width=0.2mm](\M,\G) -- (\M,\A);
              \draw[solid,line width=0.4mm,color=red] (\M,\A)-- (\O,\A);
              \draw[solid,line width=0.2mm](\O,\A)--(\O,\C);
              \draw[solid,line width=0.4mm,color=red] (\O,\C)-- (\Q,\C);
              \draw[solid,line width=0.2mm](\Q,\C)--(\Q,\E);
              \draw[solid,line width=0.4mm,color=red] (\Q,\E)-- (\S,\E);
               \draw[solid,line width=0.2mm](\S,\E)--(\S,\C);
               \draw[solid,line width=0.4mm,color=red] (\S,\C)-- (\U,\C);
               \draw (1.5cm,1.5cm) node {\scriptsize $0$};
               \draw (3.5cm,1.5cm) node {\scriptsize $0$};
               \draw (5.5cm,3.5cm) node {\scriptsize $1$};
               \draw (7.5cm,5.5cm) node {\scriptsize $2$};
               \draw (9.5cm,7.5cm) node {\scriptsize $3$};
               \draw (11.5cm,7.5cm) node {\scriptsize $3$};
               \draw (13.5cm,1.5cm) node {\scriptsize $0$};
               \draw (15.5cm,3.5cm) node {\scriptsize $1$};
               \draw (17.5cm,5.5cm) node {\scriptsize $2$};
               \draw (19.5cm,3.5cm) node {\scriptsize $1$};
               \draw (11cm,-3.5cm) node {\scriptsize $(b)$};
         \end{tikzpicture}
    \end{center}
 \caption{\label{Fig}(a) A Dyck path where each up step is labeled by the ordinate of its starting point;
and (b) its corresponding Catalan word $0012330121$.}
\label{fig3}
\end{figure}

\section{Catalan words vs. ascent sequences}
\label{sect_ascent}
An {\it ascent} in a word $\boldsymbol{w}=w_1w_2\dots w_n$ is a position $i$, $1\leq i\leq n-1$,
with $w_i<w_{i+1}$, and $\mathrm{asc}(\boldsymbol{w})$ denotes the number of ascents in $\boldsymbol{w}$.
Closely related to Catalan words are ascent sequences introduced in \cite{BCDK} and defined as: the word $\boldsymbol{w}=w_1w_2\ldots w_n$ is called an 
{\it ascent sequence} if 
$$w_1=0\mbox{ and } 0\leq w_i\leq \mathrm{asc}(w_1w_2\dots w_{i-1})+1\mbox{ for } i=2,3,\ldots ,n,$$
and $\mathcal{A}_n$ denotes the set of length $n$ ascent sequences, and $\mathcal{A}=\cup_{n\geq 0}\mathcal{A}_n$.
Similarly as for Catalan words, if $\pi$ is a pattern, then 
$\mathcal{A}_n(\pi)$ is the set of length $n$ ascent sequences avoiding $\pi$, and $\mathcal{A}(\pi)=\cup_{n\geq 0}\mathcal{A}_n(\pi)$.
Clearly, $\mathcal{C}_n=\mathcal{A}_n$ for $n\leq 3$, and
$\mathcal{C}_n\subset\mathcal{A}_n$ for $n\geq 4$, and this inclusion is strict, for instance $0102\in\mathcal{A}_4\setminus\mathcal{C}_4$.
It turns out that, for particular patterns $\pi$, $\mathcal{A}_n(\pi)$ collapses to $\mathcal{C}_n(\pi)$
for any $n$, and this behaviour where the pattern $0102$ plays a critical role
is discussed below.

\begin{Pro}
If $\boldsymbol{w}\in \mathcal{A}\setminus\mathcal{C}$, then $\boldsymbol{w}$ contains the pattern $0102$.
\label{f_0102}
\end{Pro}
\proof 
If $\boldsymbol{w}=w_1w_2\dots w_n$ is an ascent sequence which is not a Catalan word, then there is an $i$ such that $w_i\geq w_{i-1}+2$, and let $k$ be the smallest such $i$. It follows that $w_i\leq w_{i-1}+1$ for any $i$, $2\leq i\leq k-1$, or equivalently $w_1w_2\dots w_{k-1}$ is a Catalan word. Thus, if $w_i>0$, $2\leq i\leq k-1$, then each symbol less than $w_i$ occurs in the prefix $w_1w_2\dots w_{i-1}$.
We distinguish two cases: (i) $w_{k-1}$ is not the maximal symbol of the prefix $w_1w_2\dots w_{k-1}$, and (ii) otherwise.

\noindent
(i) In this case there exist $i$ and $j$, $1\leq i<j< k-1$, such that $w_j=w_{k-1}+1$ and $w_i=w_{k-1}$.
It follows that $w_iw_jw_{k-1}w_k$ is an occurrence of $0102$.

\noindent
(ii) In this case the prefix $w_1w_2\dots w_{k-1}$ has a descent (otherwise, since $\boldsymbol{w}$ is an ascent sequence, the maximal possible value for $w_k$ is $w_{k-1}+1$), and let $j$ be such a descent, that is $w_j>w_{j+1}$, $j+1<k-1$. As noticed above, the symbol $w_{j+1}$ already occurs in $w_1w_2\dots w_{j-1}$, say in position $i$. Thus, $w_iw_jw_{j+1}w_k$ is an occurrence of $0102$.
\endproof

\begin{Pro}
For $n\geq 4$ and a pattern $\pi$ the followings are equivalent
\begin{enumerate}[parsep=-0.5mm,topsep=0mm]
\item $\mathcal{A}_n(\pi)=\mathcal{C}_n(\pi)$,
\item $0102$ contains the pattern $\pi$.
\end{enumerate}
\label{A-C}
\end{Pro}
\proof `2. $\Rightarrow$ 1.' We proceed by contraposition and considering $\mathcal{C}_n(\pi)\subseteq\mathcal{A}_n(\pi)$. If $\boldsymbol{w}\in \mathcal{A}_n(\pi)\setminus \mathcal{C}_n(\pi)$, then, by Proposition \ref{f_0102}, $\boldsymbol{w}$ contains $0102$, and so $0102$ does not contain $\pi$.

\noindent 
`1. $\Rightarrow$ 2.' Again by contraposition: if $0102$ does not contain $\pi$, then at least one of the words $01023\cdots (n-2)$ or $01020^{n-4}$ belongs to $\mathcal{A}_n(\pi)$ and not to $\mathcal{C}_n(\pi)$,
and so $\mathcal{A}_n(\pi)\neq\mathcal{C}_n(\pi)$.
\endproof

Since the only patterns of length three of $0102$ are $001$, $010$, $012$ and $102$, we have the following consequence of Proposition \ref{A-C}.

\begin{Co}
For $n\geq 4$ and a pattern $\pi$ of length three, $\mathcal{A}_n(\pi)=\mathcal{C}_n(\pi)$ if and only if $\pi\in\{001,010,012,$ $102\}$.
\label{cor_asc_cat}
\end{Co}
Pattern avoidance in ascent sequences was initiated in \cite{Duncan}, and in \cite{Baxter_Pudwell_15}
ascent sequences avoiding a pair of patterns of length three are considered and exact enumeration 
for several such pairs are given.
In light of Corollary \ref{cor_asc_cat} it can happen that if a pattern of the avoided pair is one of the four specified in this corollary, then the resulting ascent sequences are Catalan words 
as well. The pairs of avoided patterns for which ascent sequences and Catalan words coincide, and for which the enumeration has already been considered in \cite{Baxter_Pudwell_15} are
highlighted in the summarizing Table \ref{table2}.
In order to keep the present article self-contained we fully consider these cases, our proofs being  alternative to those in~\cite{Baxter_Pudwell_15}.

\section{Avoiding a length two and a length three pattern}
\label{Sect_2_3}
There are three patterns of length two, namely $00$, $01$ and $10$, and we have:
\begin{Pro}
\label{Pr2_3}
The number of Catalan words avoiding a pattern of length two and 
a pattern $\pi$ of length three is given by:

\begin{tabular}{ll}
$
c_n(00,\pi)=\left\{ \begin {array}{ll}
0 & {\rm if\ } \pi=012 {\rm\ and\ } n\geq 3, \\
1 & {\rm  elsewhere;}
\end {array}
\right.
$
&
$
c_n(01,\pi)=\left\{ \begin {array}{ll}
0 & {\rm if\ } \pi=000 {\rm\ and\ } n\geq 3, \\
1 & {\rm  elsewhere;}
\end {array}
\right.
$
\\
\\
$
c_n(10,\pi)=\left\{ \begin {array}{ll}
F_n & {\rm if\ } \pi=000, \\
n & {\rm if\ } \pi\in\{001,011,012\}, \\
2^{n-1} & {\rm  elsewhere,}
\end {array}
\right.
$
\end{tabular}

\noindent
where $F_n$ is the $n${\em th} Fibonacci number defined by $F_0=F_1=1$ and 
$F_n=F_{n-1}+F_{n-2}$, $n \geq 2$.
\end{Pro}
\proof
\comm{
If $\boldsymbol{w} \in \mathcal{C}_n(00)$, then  $\boldsymbol{w}= 012\cdots(n-1)$. Thus,
$\mathcal{C}_n(00,\pi)=\{012\cdots(n-1)\}$ except when
$\pi=012$ and in this case $\mathcal{C}_n(00,\pi)=\varnothing$ for $n\geq3$, and the
counting relation for $\mathcal{C}_n(00,\pi)$ follows.\\
Similarly, if $\boldsymbol{w} \in \mathcal{C}_n(01)$, then  $\boldsymbol{w}= 00\cdots0$. Thus,
$\mathcal{C}_n(01,\pi)=\{000\cdots 0\}$ except when
$\pi=000$ and in this case $\mathcal{C}_n(01,\pi)=\varnothing$ for $n\geq3$.

Finally, a Catalan word avoids $10$ if and only if it avoids $010$. It follows that 
$c_n(10,\pi)=c_n(010,\pi)$,
which falls in  the case of avoidance of two length 3 patterns
and the corresponding proofs are given in the next section, see also Table \ref{table2}.
}
\endproof
 
\section{Trivial cases}
\label{Sect_trivial}
\subsubsection*{Superfluous patterns}
If the pattern $\tau$ contains the pattern $\sigma$,
then clearly $\mathcal{C}_n(\sigma,\tau)= \mathcal{C}_n(\sigma)$;
but this phenomenon can occur even when $\sigma$ and $\tau$ are not related by containment 
and in this case, following \cite{Baxter_Pudwell_15}, we say that $\tau$ is a {\it superfluous 
pattern} for $\sigma$.
For example, any word in $\mathcal{C}_n(012)$ is a binary word, and thus 
any pattern with at least three different symbols is a superfluous 
pattern for $012$. In Table \ref{superfluous} are listed all pairs of superfluous patterns of length three.
It is worth to mentioning that superfluousness is a transitive relation, for instance $201$ is superfluous
for $021$ which in turn is superfluous for $011$. 
So, a pattern can be superfluous for several other ones, for instance  $\tau=201$ is superfluous for each of the patterns $001,010,011,012,021, 101, 120$. Also it is easy to see that if $\tau$ is superfluous for $\sigma$,
then $\tau$ is larger lexicographically than $\sigma$.

\begin{table}[h]
\begin{center}
\begin{tabular}{c|c|c|c|c|c|c|c|c|c}
$\sigma$  & $000$ & $001$ & $010$ & $011$ & $012$ & $021$ & $101$ & $110$ & $120$\\
\hline
& $100$ & $101$ & $021$ & $021$ & $021$ & $201$ & $102$ & $210$ & $201$ \\
&       & $102$ & $10$0 & $101$ & $102$ & $210$ & $201$ & & $210$ \\
&       & $201$ & $101$ & $102$ & $120$ & & & & \\
$\tau$ &       &       & $102$ & $110$ & $201$ & & & & \\
&       &       & $110$ & $201$ & $210$ & & & & \\
&       &       & $120$ & $210$ & & & & & \\
&       &       & $201$ & & & & & & \\
&       &       & $210$ & & & & & & \\
 \end{tabular}
\end{center}
\caption{\label{superfluous}
In each column, each pattern $\tau$ is superfluous for the pattern $\sigma$.}
\end{table}
\subsubsection*{Ultimately constant sequences}
It can happen that the number of Catalan words avoiding a pair of length $3$
patterns is constant for enough long words. The only two such cases are 
given below.
\begin{Pro}\label{UC}$ $
\begin{tabular}{lcl}
$
c_n(000,011)=\left\{ \begin {array}{ll}
1 & {\rm if\ } n=1, \\
2 & {\rm if\ } n=2, \\
3 & {\rm if\ } n\geq 3;
\end {array}
\right.
$
&
and
&
$
c_n(000,012)=\left\{ \begin {array}{ll}
1 & {\rm if\ } n=1, \\
2 & {\rm if\ } n=2, \\
3 & {\rm if\ } n=3{\rm\ or\ } n=4,\\
0 & {\rm if\ } n\geq 5.
\end {array}
\right.
$
\end{tabular}
\end{Pro}
\proof
\comm{
If $n\geq 3$, then $\mathcal{C}_n(000,011)=\{0\boldsymbol{x},\boldsymbol{x}0,\boldsymbol{x}(n-1)\}$ where $\boldsymbol{x}$ is the word $01\cdots (n-2)$, and the first point follows.\\
If a Catalan word avoids $012$, then it is a binary word. In addition 
if its length is larger than $4$ it necessarily contains three identical entries, and so 
$\mathcal{C}_n(000,012)=\varnothing$ for $n\geq 5$. Considering the initial values of $c_n(000,012)$ the second point follows.
}
\endproof
\subsubsection*{Counting sequence $n$}
\begin{Pro} 
\label{seq_n}
If $\pi$ is one of the pairs of patterns 
$\{001,010\}$, $\{001,011\}$, $\{001,012\}$, $\{010,011\}$, $\{010,012\}$ or $\{011,012\}$,
then $c_n(\pi)=n$.
\end{Pro}
\proof
\comm{
The proof is based on the easy to understand description given below 
for the corresponding sets of Catalan words:

\medskip

\hspace{-4mm}$
\begin {array}{c|c||c|c}
\pi & \mathcal{C}_n(\pi) & \pi & \mathcal{C}_n(\pi)\\ \hline
\{001,010\} & \{012\cdots jj\cdots j\,:\, 0\leq j\leq n-1\} 
       & \{010,011\} & \{0^j 12\cdots (n-j)\,:\, 1\leq j\leq n\}\\
\{001,011\} & \{012\cdots j0\cdots 0\,:\, 0\leq j\leq n-1\} 
       & \{010,012\} &\{0^{j+1}1^{n-j-1}\,:\, 0\leq j\leq n-1\}\\
\{001,012\} & \{01^j0^{n-j-1}\,:\, 0\leq j\leq n-1\}        
       & \{011,012\} & \{0^n\}\cup \{0^j10^{n-j-1}\,:\, 1\leq j\leq n-1\}\\ 
\end{array}
$
}
\endproof
\section{Counting via recurrence}
\label{Sect_recure}

\subsubsection*{Counting sequence $2(n-1)$}
\begin{Pro} \label{2n1}
If $\pi=\{011,100\}$ or $\pi=\{011,120\}$, then $c_n(\pi)=2(n-1)$ for $n\geq2$
(Sequence $\href{https://oeis.org/A005843 }{\tt A005843}$ in \cite{OEIS}).\\
\end{Pro}
\proof
\comm{
If $\pi=\{011,100\}$ and $\boldsymbol{w}\in\mathcal{C}_n(\pi)$, $n\geq2$, then either
\begin{enumerate}[parsep=-0.5mm,topsep=0mm]
\item[$-$] $\boldsymbol{w}=0^k\boldsymbol{u} $, $0\leq k\leq n-1$, with $\boldsymbol{u} $ a s.i. Catalan word
       (of length at least one), or 
\item[$-$] $\boldsymbol{w}=0^k\boldsymbol{u} 0$, $0\leq k\leq n-3$, with $\boldsymbol{u} $ a s.i.  Catalan
        word (of length at least two).
\end{enumerate}
In the first case there are $n$ possibilities for $\boldsymbol{w}$ and $n-2$ possibilities in the second case, and the result holds.
\noindent
If $\pi=\{011,120\}$ and $\boldsymbol{w}\in\mathcal{C}_n(\pi)$, $n\geq2$, then either
\begin{enumerate}[parsep=-0.5mm,topsep=0mm]
\item[$-$] $\boldsymbol{w}=0^k\boldsymbol{u}$, $0\leq k\leq n-1$, with 
      $\boldsymbol{u}$ a {s.i.} Catalan word, or 
\item[$-$] $\boldsymbol{w}=0^k10^{n-k-1}$, $1\leq k\leq n-2$,
\end{enumerate}
and as previously the result holds.
}
\endproof

\subsubsection*{Sequences involving $2^n$}
\begin{Pro}
\label{2_pw_n}
If $\pi=\{000,101\}$, then $c_n(\pi)=2^{n-1}$ for $n\geq 1$ ($\href{https://oeis.org/A000079 }{\tt A000079}$ in \cite{OEIS}).
\end{Pro}
\proof
\comm{
If a Catalan word avoids $101$, then it is unimodal (that is, it can be written not necessarily 
in a unique way as 
$\boldsymbol{uv}$ with $\boldsymbol{u}$ a weakly decreasing
 and $\boldsymbol{v}$ weakly decreasing word). In addition, if the word avoids $000$, then its maximal value occurs at most twice, and when it occurs twice this happens in consecutive positions.\\
We denote by $\mathcal{D}_n$ the subset of words in $\mathcal{C}_n(\pi)$ where the maximal entry occurs once and by $\mathcal{E}_n$ that where it occurs twice, $d_n=|\mathcal{D}_n|$, $e_n=|\mathcal{E}_n|$, and
$c_n(\pi)=d_n + e_n$.
Any word $\boldsymbol{w}=w_1 \cdots mm\cdots w_{n-1} \in \mathcal{E}_{n-1}$ with  its maximal value $m$ occurring twice can be extended into a word in $\mathcal{D}_n$ by one of the transformations:
\begin{itemize}
\item[] $\boldsymbol{w}\mapsto w_1 \cdots m(m+1)m\cdots w_{n-1}$, and
\item[] $\boldsymbol{w}\mapsto w_1 \cdots mm(m+1)\cdots w_{n-1}$,
 \end{itemize}
and any word $\boldsymbol{w}=w_1 \cdots m\cdots w_{n-1} \in \mathcal{D}_{n-1}$ with its maximal value $m$ occurring once can be extended into a word in $\mathcal{D}_n$ by:
\begin{itemize}
\item[] $\boldsymbol{w}\mapsto w_1 \cdots m(m+1)\cdots w_{n-1}$.
\end{itemize}
Conversely, any word in $\mathcal{D}_n$, $n\geq 2$, can uniquely be obtained  from a word in $\mathcal{D}_{n-1}$ or in $\mathcal{E}_{n-1}$ by reversing one of the transformations above, so
$$
d_n=2\cdot e_{n-1}+d_{n-1},
$$
for $n\geq2$.\\
Reasoning in a similar way we have
$$
e_n=2\cdot e_{n-2}+d_{n-2},
$$
Thus, for $n\geq3$, $e_n=d_{n-1}$, and finally
$$
c_n(\pi)=d_n+e_n=2\cdot (d_{n-1} + e_{n-1})=2\cdot c_{n-1}(\pi),
$$
and with the initial conditions $c_1(\pi)=1$ and $c_2(\pi)=2$, the result follows.
}
\endproof

\begin{Pro}
\label{2nmp1}
If $\pi=\{101,210\}$ or $\pi=\{102,120\}$, then
$c_n(\pi)=(n-1)\cdot 2^{n-2}+1$ for $n\geq2$
($\href{https://oeis.org/A005183 }{\tt A005183}$ in \cite{OEIS}).
\end{Pro}
\proof
\comm{
If $\pi=\{101,210\}$ and $\boldsymbol{w}\in \mathcal{C}_n(\pi)$, $n\geq3$, then either
\begin{enumerate}[parsep=-0.5mm,topsep=0mm]
\item[$-$] $\boldsymbol{w}=0\boldsymbol{u}$ with $\boldsymbol{u} \in \mathcal{C}_{n-1}(\pi)$, or
\item[$-$] $\boldsymbol{w}=0(\boldsymbol{u}+1)$ with $\boldsymbol{u} \in \mathcal{C}_{n-1}(\pi)$, or
\item[$-$] $\boldsymbol{w}=0(\boldsymbol{u}+1)0^{n-k-1}$
with $\boldsymbol{u} $ a {w.i.} Catalan word of length 
$k$, $1\leq k\leq n-2$.
\end{enumerate}
The number of words in each of the first two cases is $c_{n-1}(\pi)$.
The number of length $k$ w.i. Catalan words is $2^{k-1}$, so the number of words in the last case is 
$$
\sum^{n-2}_{k=1} 2^{k-1}= \sum^{n-3}_{k=0} 2^{k}= 2^{n-2}-1.
$$
Thus, $c_n(\pi)=2c_{n-1}(\pi)+ 2^{n-2}-1$, and after calculation the result holds.
\medskip

\noindent
If $\pi=\{102,120\}$ and $\boldsymbol{w}\in \mathcal{C}_n(\pi)$, $n\geq3$, then either
\begin{enumerate}[parsep=-0.5mm,topsep=0mm]
\item[$-$] $\boldsymbol{w}       =0\boldsymbol{u}$ with $\boldsymbol{u} \in \mathcal{C}_{n-1}(\pi)$, or
\item[$-$] $\boldsymbol{w}       =0(\boldsymbol{u}+1)$ with $\boldsymbol{u} \in \mathcal{C}_{n-1}(\pi)$, or
\item[$-$] $\boldsymbol{w}=01\boldsymbol{u}$ with $\boldsymbol{u} $ a length $(n-2)$ binary word other
than $11\cdots1$.
\end{enumerate}
The number of words in each of the first two cases is $c_{n-1}(\pi)$, and the number of words in the 
last case is $2^{n-2}-1$.
So $c_n(\pi)=2c_{n-1}(\pi)+ 2^{n-2}-1$, and again the result holds. 
}
\endproof

\begin{Pro}
\label{2nmn}
If  $\pi$ is one of the pairs of patterns $\{021,100\}$, $\{021,101\}$,
$\{101,110\}$ or $\{101,120\}$,
then $c_n(\pi)= 2^n-n$ for $n\geq 0$ ($\href{https://oeis.org/A000325}{\tt A000325}$ in $\cite{OEIS}$).
\end{Pro}
\proof
\comm{

\noindent
If $\pi=\{021,100\}$ and $\boldsymbol{w}\in\mathcal{C}_n(\pi)$, $n\geq4$, then either
\begin{enumerate}[parsep=-0.5mm,topsep=0mm]
\item[$-$] $\boldsymbol{w}$ is a w.i. Catalan word, or
\item[$-$] $\boldsymbol{w}=\boldsymbol{u}0$ with $\boldsymbol{u}$ a w.i. Catalan word of length $(n-1)$ other than $00\cdots 0$, or
\item[$-$] $\boldsymbol{w}=0\boldsymbol{v}0(\boldsymbol{u}+1)$ where
    $\boldsymbol{v}$ is w.i. binary word ending by $1$ and 
    $\boldsymbol{u}$ is a w.i. Catalan word of length $k$, $1\leq k \leq n-3$.
\end{enumerate}
\noindent
In the first case the number of words $\boldsymbol{w}$ is $2^{n-1}$ and in the 
second case the number of words $\boldsymbol{w}$ is $2^{n-2}-1$. In last case the number of words $\boldsymbol{w}$ is
$$
\sum^{n-3}_{k=1}(n-k-2)\cdot 2^{k-1}=2^{n-2}-(n-1).
$$
Combining these cases and considering the initial values of $c_n(\pi)$
the result holds.

\medskip

\noindent
If $\pi=\{021,101\}$ and $\boldsymbol{w}       \in \mathcal{C}_n(\pi)$, $n\geq2$, then either
\begin{enumerate}[parsep=-0.5mm,topsep=0mm]
\item[$-$] $\boldsymbol{w}       =\boldsymbol{u}0$ with $\boldsymbol{u} \in \mathcal{C}_{n-1}(\pi) $, or
\item[$-$] $\boldsymbol{w}       =0^{n-k}(\boldsymbol{u}+1)$ with $\boldsymbol{u} $ a w.i. Catalan word of length $k$, $1\leq k\leq n-1$.
\end{enumerate}
The number of words in the first case is $c_{n-1}(\pi)$ and the number of words in the second
case is $\sum_{k=1}^{n-1}2^{k-1}=2^{n-1}-1$.
So $c_n(\pi)=c_{n-1}(\pi)+ 2^{n-1}-1$ and after calculation the result follows.

\medskip

\noindent
If $\pi=\{101,110\}$ and $\boldsymbol{w}\in \mathcal{C}_n(\pi)$, $n\geq 3$, then either
\begin{enumerate}[parsep=-0.5mm,topsep=0mm]
\item[$-$] $\boldsymbol{w}=0\boldsymbol{u}$ with $\boldsymbol{u} \in \mathcal{C}_{n-1}(\pi)$, or
\item[$-$] $\boldsymbol{w}=0(\boldsymbol{u}+1)$ with $\boldsymbol{u} \in \mathcal{C}_{n-1}(\pi)$, or
\item[$-$] $\boldsymbol{w}=\boldsymbol{u}0^{n-k}$ with $\boldsymbol{u}$ a s.i. Catalan word of length 
$k$, $2\leq k \leq n-1$.
\end{enumerate}
The number of words $\boldsymbol{w}$ in each of the first 
two cases is $c_{n-1}(\pi)$. 
For the last case, for each $k$ there is exactly one word $\boldsymbol{w}$,
so their number is $(n-2)$ in this case.
So $c_n(\pi)=2c_{n-1}(\pi)+n-2$ and after calculation the result holds.

\medskip

\noindent
If $\pi=\{101,120\}$ and $\boldsymbol{w}\in \mathcal{C}_n(\pi)$, $n\geq 3$, then either
\begin{enumerate}[parsep=-0.5mm,topsep=0mm]
\item[$-$] $\boldsymbol{w}=0\boldsymbol{u}$ with $\boldsymbol{u}\in\mathcal{C}_{n-1}(\pi)$, or
\item[$-$] $\boldsymbol{w}=0(\boldsymbol{u}+1)$ with $\boldsymbol{u}\in\mathcal{C}_{n-1}(\pi)$, or 
\item[$-$] $\boldsymbol{w}=01^k0^{n-k-1}$, $1\leq k\leq n-2$.
\end{enumerate}
So, again $c_n(\pi)=2c_{n-1}(\pi)+n-2$.
}
\endproof

\begin{Pro}
\label{np22n}
If  $\pi=\{021,120\}$, then
$c_n(\pi)=(n+2)\cdot 2^{n-3}$ for $n\geq3$ ($\href{https://oeis.org/A045623}{\tt A045623}$ in $\cite{OEIS}$).
\end{Pro}
\proof
\comm{
If  $\boldsymbol{w}       \in \mathcal{C}_n(\pi)$, $n\geq4$, then either
\begin{enumerate}[parsep=-0.5mm,topsep=0mm]
    \item[$-$] $\boldsymbol{w}=0\boldsymbol{u}0$ with $\boldsymbol{u}$ a length $(n-2)$ binary word, or
    \item[$-$]$\boldsymbol{w} =0(\boldsymbol{u}+1)$ with $\boldsymbol{u}$ a length $(n-1)$ w.i. Catalan
     word, or
    \item[$-$]$\boldsymbol{w}       =0\boldsymbol{u}0(\boldsymbol{v}+1)$ with $\boldsymbol{u}$ a binary word and $\boldsymbol{v}$ w.i. Catalan word.
\end{enumerate}
The number of words in each of the first two cases is $2^{n-2}$ and the number of words in the last case is
$$
\sum_{k=1}^{n-2}2^{n-k-2}2^{k-1}=(n-2)2^{n-3},
$$
and so $ c_n(\pi)=2^{n-1}+(n-2)2^{n-3}$, which gives the desired result.
}
\endproof

If a Catalan word avoids both $102$ and $110$, then it has at most one descent.
In the second part of the proof of the next proposition we need the following 
technical lemma which gives the number of Catalan words in $\mathcal{C}_n(102,110)$ 
with one descent and avoiding the pattern $00$ before the descent
(note that in this case avoiding $00$ is equivalent to avoiding equal consecutive entries).
The set of these words is empty for $n\leq 2$,
it is the single word set $\{010\}$ for $n=3$, and $\{0100,0101,0120,0121\}$ for $n=4$.

\begin{Le}\label{le1}
Let $\mathcal{D}_n$ be the set of words in $\mathcal{C}_n(102,110)$ having one descent and
avoiding $00$ before the descent.
Then $|\mathcal{D}_n|=\frac{n}{6}(n-1)(n-2)$.
\end{Le}
\proof
\comm{
A word belongs to $\mathcal{D}_n$, $n\geq 3$, if and only if it can be written as
$$
012\ldots (k-1)a(\boldsymbol{u}+a),
$$
with $2\leq k\leq n-1$ ($k$ is the position of the unique descent in the word), 
$a\in\{0,1,\ldots k-2\}$, and $\boldsymbol{u}$ is a length $(n-k-1)$ w.i. Catalan  word over 
$\{0,1\}$.
For each choice of $k$, there are $k-1$ choices for $a$, and for each choice for $a$
there are $n-k$ choices for $\boldsymbol{u}$. It follows that 
$$
|\mathcal{D}_n|=\sum_{k=2}^{n-1}(k-1)\cdot (n-k),
$$
and after calculation the statement holds.
}
\endproof

\begin{Pro}
\label{32n}
If $\pi=\{021,102\}$ or $\pi=\{102,110\}$, then $c_n(\pi)=3\cdot 2^{n-1}-\frac{1}{2}(n+1)(n+2)+n$ for $n\geq1$
($\href{https://oeis.org/A116702}{\tt A116702}$ in $\cite{OEIS}$). \\
\end{Pro}
\proof
\comm{
If $\pi=\{021,102\}$ and $\boldsymbol{w}\in \mathcal{C}_n(\pi)$, $n\geq 3$, then either
\begin{enumerate}[parsep=-0.5mm,topsep=0mm]
\item[$-$] $\boldsymbol{w}=0\boldsymbol{u}$ with $\boldsymbol{u}$ a length $(n-1)$ binary word, or
\item[$-$] $\boldsymbol{w}=0\cdots0(\boldsymbol{u}+1)0\cdots0$ with $\boldsymbol{u}$ a length $k$, 
$2\leq k\leq n-1$, w.i. Catalan word other than $00\cdots 0$ and $\boldsymbol{w}$ beginning by at least one $0$.
\end{enumerate}
The number of words in the first case is $2^{n-1}$ and the number of those in the second case is
$$
\sum_{k=2}^{n-1}(2^{k-1}-1)\cdot(n-k)=2\cdot 2^{n-1}-\frac{1}{2}(n+1)(n+2)+n,
$$
and combining the two cases the result holds.

\medskip
\noindent

If $\pi=\{102,110\}$ and $\boldsymbol{w}\in\mathcal{C}_n(\pi)$, then either
\begin{enumerate}[parsep=-0.5mm,topsep=0mm]
\item[$-$] $\boldsymbol{w}$ is a w.i. Catalan word, the number of such words is $2^{n-1}$, or
\item[$-$] $\boldsymbol{w}\in\mathcal{D}_n$, with $\mathcal{D}_n$ defined in Lemma \ref{le1}, or
\item[$-$] $\boldsymbol{w}=\boldsymbol{u}(\boldsymbol{v}+m)$, where $\boldsymbol{u}$ is a w.i. Catalan word of length $k$, $1\leq k\leq n-3$, 
$m$ is the maximal (rightmost) entry of $\boldsymbol{u}$, and $\boldsymbol{v}$ is a word belonging to $\mathcal{D}_{n-k}$. 
\end{enumerate}
Indeed, the first case corresponds to words with no descents, 
the second one to those with a descent and no occurrences of $00$ before the descent, and the third one 
to those with both descent and occurrences of $00$ before the descent (the rightmost 
such occurrence is in positions $k$ and $k+1$).
By Lemma \ref{le1}, the number of words in the third case is 
$$
\sum_{k=1}^{n-3}2^{k-1}\cdot (n-k)\cdot
\left(\frac{1}{6}(n-k)^2-\frac{1}{2}(n-k)+\frac{1}{3}\right)=
2^n-\frac{1}{6}(n+1)(n^2-n+6).
$$
Finally, combining the three previous cases the desired result holds. 
}
\endproof
 
\subsubsection*{Sequences involving binomial coefficients}

In this part we use the notation
$$
ab \cdots  c\underset{\underset{i}{\uparrow}}{d}e \cdots f
$$
to denote that the entry $d$ is in position $i$ in the word 
$ab \cdots  cde \cdots f$.

\begin{Pro}
\label{C3}
If  $\pi=\{001,210\}$, then $c_n(\pi)={n\choose 3}+n$ for $n\geq3$ ($\href{https://oeis.org/A000125}{\tt A000125}$ in $\cite{OEIS}$).
\end{Pro}
\proof
\comm{
If $\boldsymbol{w}\in\mathcal{C}_n(\pi)$, then it has at most one descent.

\noindent
If $\boldsymbol{w}$ has no descents, then it has the form
$\boldsymbol{w}=01\cdots (m-1)m\cdots m$ and there are $n$ such words.

\noindent
If $\boldsymbol{w}$ has one descent, then it has the form
$\boldsymbol{w}=01\cdots (m-1)m\cdots mk\cdots k$ with $0\leq k<m<n-1$.
It follows that there is a bijection between the family of $3$-element subsets of $\{0,1,\dots,n-1\}$ and 
the words in $\mathcal{C}_n(\pi)$ with one descent:
$$
\{k,m,\ell\}\mapsto 0123 \cdots (m-1)m \cdots 
\underset{\underset{\ell}{\uparrow}}{m}k \cdots k.
$$
Combining the two cases we have $ c_n(\pi)= {n\choose 3}+n$.
}
\endproof

\begin{Pro}
\label{C2}
If  $\pi$ is one of the pairs of patterns $\{001, 021\}$, $\{001,110\}$, $\{001,120\}$,
$\{012,100\}$, $\{012,101\}$ or $\{012,110\}$, then
$c_n(\pi)={n\choose 2}+1$ for $n\geq2$ ($\href{https://oeis.org/A000124}{\tt A000124}$ in $\cite{OEIS}$).
\end{Pro}
\proof
\comm{
In any of the six cases for $\pi$, the set $\mathcal{C}_n(\pi)$ is in bijection
with the family $S$ of subsets of $\{2,\dots,n\}$ with at most two elements.
We give below explicit definitions for such bijections,
where the empty set is mapped to $0\cdots0\in\mathcal{C}_n(\pi)$ by each of them.

\medskip

\noindent
If $\pi=\{001,021\}$ and $\boldsymbol{w}\in\mathcal{C}_n(\pi)$, then either
$\boldsymbol{w}=00\cdots 0$ or for some $m\geq 1$,
$\boldsymbol{w}=0123\cdots (m-1)m^{s}$ with $s\geq 1$
or $\boldsymbol{w}=0123\cdots (m-1)m^{s}0^{t}$ with  $s\geq 1$, $t\geq 1$,
and the desired bijection $S\rightarrow\mathcal{C}_{n}(001,021)$ is
\begin{itemize}[parsep=-2mm,topsep=0.0mm]
\item[] 
$\{k\}\mapsto 0123\cdots (m-1)\underset{\underset{k=m+1}{\uparrow}}{m} \cdots m$;
\item[] 
$\{k,j\}\mapsto0123 \cdots (m-1)\underset{\underset{k=m+1}{\uparrow}}{m}\cdots m 
\underset{\underset{j}{\uparrow}}{0}\cdots 0$.
\end{itemize}
\noindent
If $\pi=\{001,110\}$ and $\boldsymbol{w}\in\mathcal{C}_n(\pi)$, then either
$\boldsymbol{w}=00\cdots 0$
or
$\boldsymbol{w}=0123\cdots (m-1)m\cdots m$ with $m$ the maximal entry of $\boldsymbol{w}$, or
$\boldsymbol{w}=0123\cdots (m-1)m \ell  \cdots\ell$ with $\ell<m$, and the desired bijection 
$S\rightarrow\mathcal{C}_{n}(001,110)$ is
\begin{itemize}[parsep=-2mm,topsep=0.0mm]
\item[] 
$\{k\}\mapsto  0123\cdots (m-1)\underset{\underset{k=m+1}{\uparrow}}{m} \cdots m$;
\item[] 
$\{k,j\}\mapsto 0123 \cdots (j-2)(k-2)\cdots (k-2)$.
\vspace{0.1cm}
\end{itemize}

\noindent
If $\pi=\{001,120\}$ and $\boldsymbol{w}\in\mathcal{C}_n(\pi)$, then either 
$\boldsymbol{w}=00\cdots 0$,
or for some $m\geq 1$,
$\boldsymbol{w}=0123\cdots (m-1)m^{s}$ with $s\geq 1$
or
$
\boldsymbol{w}=0123\cdots (m-1)m^{s}(m-1)^{t}$ with $s,t\geq 1$,
and the desired bijection $S \rightarrow\mathcal{C}_{n}(001,120) $ is
\begin{itemize}[parsep=-2mm,topsep=0.0mm]
\item[] 
$\{k\}\mapsto 0123\cdots (m-1) \underset{\underset{k=m+1}{\uparrow}}{m}\cdots m$;
\item[]
$\{k,j\}\mapsto0123 \cdots (m-1)\underset{\underset{k=m+1}{\uparrow}}{m} \cdots m 
\underset{\underset{j}{\uparrow}}{(m-1)}\cdots (m-1)$.
\end{itemize}

For the next three cases we need the following observation:
a Catalan word avoids  $012$ if and only if it is a binary word (over $\{0,1\}$) beginning by a $0$.
\medskip

\noindent
If $\pi=\{012,100\}$ and $\boldsymbol{w}\in\mathcal{C}_n(\pi)$, then either
$\boldsymbol{w}=0^{s}1^{t}$ with $s\geq 1$, $t\geq 0$ or
$\boldsymbol{w}=0^{s}1^{t}01^{r}$ with $s,t\geq 1$, $r\geq0$,
and the desired bijection $S\rightarrow  \mathcal{C}_{n}(012,100) $ is
\begin{itemize}[parsep=-2mm,topsep=-0.0mm]
\item[] 
$\{k\} \mapsto 0\cdots0\underset{\underset{k}{\uparrow}}{1} \cdots 1$;
\item[] $\{k,j\}\mapsto 0\cdots0\underset{\underset{k}{\uparrow}}{1} \cdots 1
\underset{\underset{j}{\uparrow}}{0}1\cdots 1$. 
\end{itemize}
\noindent
If $\pi=\{012,101\}$ and $\boldsymbol{w}\in\mathcal{C}_n(\pi)$, then either 
$\boldsymbol{w}=0^{s}1^{t}$ with $s\geq 1$, $t\geq 0$
or
$\boldsymbol{w}=0^{s}1^{t}0^{r}$ with $s, t\geq 1$, $r\geq 1$, 
and the desired bijection $S\rightarrow  \mathcal{C}_{n}(012,101) $ is
\begin{itemize}[parsep=-2mm,topsep=0.0mm]
\item[] 
$\{k\}\mapsto 0\cdots0\underset{\underset{k}{\uparrow}}{1} \cdots 1$;
\item[] 
$\{k,j\}\mapsto 0\cdots0\underset{\underset{k}{\uparrow}}{1} \cdots 1
\underset{\underset{j}{\uparrow}}{0} \cdots0$.
\end{itemize}

\noindent
If $\pi=\{012,110\}$ and $\boldsymbol{w}\in\mathcal{C}_n(\pi)$, then either
$\boldsymbol{w}=0^{s}1^{t}$ with $s\geq 1$, $t\geq 0$
or
$\boldsymbol{w}=0^{s}10^{t}1^{r}$ with $s, t\geq 1$, $r\geq0$, and the 
desired bijection $S \rightarrow  \mathcal{C}_{n}(012,110)$ is
\begin{itemize}[parsep=-2mm,topsep=0.0mm]
\item[] 
$\{k\}\mapsto 0\cdots0\underset{\underset{k}{\uparrow}}{1}\cdots 1$;
\item[] 
$\{k,j\}\mapsto  0\cdots 0\underset{\underset{k}{\uparrow}}{1}
0\cdots \underset{\underset{j}{\uparrow}}{0}1\cdots 1$.
\end{itemize}

}
\endproof
\subsubsection*{Sequences involving Fibonacci(-like) numbers}
As in Proposition \ref{Pr2_3}, we consider the sequence of Fibonacci numbers $(F_n)_{n\geq 0}$ defined as $F_0=1$, $F_1=1$ and $F_n=F_{n-1} + F_{n-2}$ for $n\geq2$.

\begin{Pro}
\label{Fibo}
If $\pi=\{000,001\}$ or $\pi=\{000,010\}$, then
$c_n(\pi)=F_n$ for $n\geq 0$ ($\href{https://oeis.org/A00045}{\tt A00045}$ in $\cite{OEIS}$).
\end{Pro}
\proof
\comm{
For $\pi=\{000,001\}$ the proof is up to a certain point similar to that of Proposition \ref{2_pw_n}.
A word belonging to $\mathcal{C}_n(\pi)$ is unimodal and its maximal entry occurs once or twice
in consecutive positions. Let $\mathcal{D}_n$ denote the subset of words in $\mathcal{C}_n(\pi)$ where 
the maximal entry occurs once and $\mathcal{E}_n$ denote that where it occurs twice.
If $\boldsymbol{w}\in\mathcal{C}_n(\pi)$ has its maximal entry $m$, then
the insertion of $(m+1)$ after the leftmost occurrence of $m$ in $\boldsymbol{w}$
produces a word in $\mathcal{D}_{n+1}$, and the insertion of $(m+1)(m+1)$ 
produces a word in $\mathcal{E}_{n+2}$. It is easy to see that these transformations induce
a bijection between $\mathcal{C}_n(\pi)$ and $\mathcal{D}_{n+1}$, and between $\mathcal{C}_n(\pi)$ and $\mathcal{E}_{n+2}$,
and thus between $\mathcal{C}_{n-2}(\pi)\cup \mathcal{C}_{n-1}(\pi)$ and $\mathcal{C}_n(\pi)$.
It follows that $c_n(\pi)$ satisfies a Fibonacci-like recurrence, and by considering the initial 
values for $c_n(\pi)$ the result holds.

\medskip

\noindent
For $\pi=\{000,010\}$, a word $\boldsymbol{w}\in {C_n(\pi)}$ is characterized by:
$\boldsymbol{w}$ is w.i. and $\boldsymbol{w}$ does not have three consecutive equal entries. So $\boldsymbol{w}$ can be represented by the binary word $b_1b_2\dots b_{n-1}$ with no two consecutive $1$s where $b_i=1$ iff $w_i=w_{i-1}$.
This representation is a bijection between  $\mathcal{C}_n(\pi)$ and the set of binary words of length $(n-1)$ without two consecutive $1$s, which cardinality is the Fibonacci number,
see for instance \cite{Vaj_Fibo}.
}
\endproof

\begin{Pro}
\label{Fibom1}
If $\pi=\{001,100\}$, then $c_n(\pi)=F_{n+1}-1$ for $n\geq 1$.
\end{Pro}
\proof 
\comm{
If  $\boldsymbol{w}\in \mathcal{C}_n(\pi)$, $n\geq 3$, then either
\begin{enumerate}[parsep=-0.5mm,topsep=0mm]
\item[$-$] $\boldsymbol{w}=0\dotsb 0$, or
\item[$-$] $\boldsymbol{w}=0(\boldsymbol{u}+1)$ with $\boldsymbol{u} \in \mathcal{C}_{n-1}(\pi) $, or
\item[$-$] $\boldsymbol{w}=0(\boldsymbol{u}+1)0$ with $\boldsymbol{u} \in \mathcal{C}_{n-2}(\pi)$. 
\end{enumerate}
So, $c_n(\pi)$ satisfies the recurrence $c_n(\pi) =c_{n-1}(\pi)+c_{n-2}(\pi)+1$ for $n\geq 3$, and solving it
we have the desired result.
}
\endproof

In the next proposition we will make use of the following relation satisfied by the 
even index Fibonacci numbers: $F_{2n}=F_{2n-2}+\sum_{i=0}^{n-1}F_{2i}$ for $n\geq 1$.

\begin{Pro}
\label{Fibo_bis}
If $\pi=\{100,201\}$, then
$c_n(\pi)=F_{2n-2}$ for $n\geq1$ ($\href{https://oeis.org/A001519}{\tt A001519}$ in $\cite{OEIS}$).
\end{Pro}
\proof
\comm{
If $\boldsymbol{w}\in \mathcal{C}_n(\pi)$, $n\geq 3$, then either
\begin{enumerate}[parsep=-0.5mm,topsep=0mm]
\item[$-$] $\boldsymbol{w}=0\boldsymbol{u}$ with $\boldsymbol{u} \in \mathcal{C}_{n-1}(\pi)$, or
\item[$-$] $\boldsymbol{w}=0(\boldsymbol{u}+1)$ with $\boldsymbol{u} \in \mathcal{C}_{n-1}(\pi)$, or
\item[$-$] $\boldsymbol{w}=0(\boldsymbol{u}+1)0$ with $\boldsymbol{u} \in \mathcal{C}_{n-2}(\pi)$, or
\item[$-$]
$\boldsymbol{w}=01^{n-k-2}0(\boldsymbol{u}+1)$ with $\boldsymbol{u} \in \mathcal{C}_{k}(\pi)$ for some $k$, $1\leq k \leq n-3$.
\end{enumerate}
In both of the first two cases the numbers of words $\boldsymbol{w}$ is $c_{n-1}(\pi)$ and in the third case, this number is $c_{n-2}(\pi)$. In the last case, the number of words $\boldsymbol{w}$ is
$
\sum^{n-3}_{k=1} c_{k}(\pi).
$
So, $c_n(\pi)=2c_{n-1}(\pi)+c_{n-2}(\pi)+ \sum_{k=1}^{n-3}c_k(\pi)=c_{n-1}(\pi)+ \sum_{k=1}^{n-1}c_k(\pi)$, and with the initial conditions we have $c_n(\pi)= F_{2n-2}$.
}
\endproof

The sequence of Pell numbers $(p_n)_{n\geq 0}$ is defined as 
$p_0=0$, $p_1=1$ and $p_n=2p_{n-1} + p_{n-2}$ for $n\geq2$.

\begin{Pro}
\label{Pell}
If $\pi=\{100,101\}$, then $c_n(\pi)$ is  the $n${\em th} Pell number for $n\geq 1$ ($\href{https://oeis.org/A000129}{\tt A000129}$ in $\cite{OEIS}$).  
\end{Pro}
\proof
\comm{
If $\boldsymbol{w}\in \mathcal{C}_n(\pi)$, $n\geq 2$, then either
\begin{enumerate}[parsep=-0.5mm,topsep=0mm]
 \item[$-$] $\boldsymbol{w}=0\boldsymbol{u}$ with $\boldsymbol{u}\in\mathcal{C}_{n-1}(\pi)$, or
\item[$-$] $\boldsymbol{w}=0(\boldsymbol{u}+1)$ with $\boldsymbol{u}\in\mathcal{C}_{n-1}(\pi)$, or
 \item[$-$] $\boldsymbol{w}       =0(\boldsymbol{u}+1)0$ with $\boldsymbol{u} \in \mathcal{C}_{n-2}(\pi)$. 
\end{enumerate}
In both of the first two cases the numbers of words $\boldsymbol{w}$
is $c_{n-1}(\pi)$ and it is $c_{n-2}(\pi)$ in the last case. 
So, $c_n(\pi)$ satisfies Pell numbers recurrence and considering its 
initial values the statement holds.
}
\endproof

\section{Counting via generating function}
\label{Sect_gen}

Here we give bivariate generating functions $C_\pi(x,y)$ 
where the coefficient of $x^ny^k$ is the number of Catalan words of length $n$ 
having $k$ descents and avoiding $\pi$, 
for each of the remaining pairs $\pi$ of patterns of length~$3$. 
Plugging $y=1$ in $C_\pi(x,y)$ we obtain $C_\pi(x)=C_\pi(x,1)$
where  the coefficient of $x^n$ is the number of Catalan words of length $n$ avoiding $\pi$. 
All the obtained enumerating sequences are not yet recorded in \cite{OEIS}, 
except that  for: $\pi=\{100,120\}$ and for $\pi=\{110,120\}$ (see
Corollary \ref{Co_100_120}) and presumably for $\pi=\{100,210\}$ (see Corollary \ref{Co_100_210}). 
In almost all the proofs of the next propositions the desired 
generating function is the solution of a functional equation satisfied by it.

\begin{Pro}\label{000_021}
If $\pi=\{000,021\}$, then
\begin{equation*}
    C_\pi(x,y)= - \frac{x^4y+x^2y+1}{x^2+x-1}.
\end{equation*}
\end{Pro}
\proof
\comm{
Here we need the generating function for the Fibonacci numbers 
$C_{000,010}(x)=\frac{1}{1-x-x^2}$ for the set in Proposition \ref{Fibo}.
Note that words in  $\mathcal{C}(000,010)$ have no descents.

\noindent
Let $\boldsymbol{w}$ be a non-empty word in $\mathcal{C}(\pi)$. Then $\boldsymbol{w}$ has one of the following forms:
\begin{itemize}[parsep=-0.5mm,topsep=0mm]
\item[$-$] $\boldsymbol{w}=0(\boldsymbol{u}+1$) where 
      $\boldsymbol{u} \in \mathcal{C}(000,010) $; the generating function for these words is 
      $x\cdot \frac{1}{1-x-x^2}$,
\item[$-$] $\boldsymbol{w}=00(\boldsymbol{u}+1)$ where $\boldsymbol{u} \in C(000,010)$;
     the generating function for these words is  $x^2\cdot\frac{1}{1-x-x^2}$,
\item[$-$] $\boldsymbol{w}  =0 (\boldsymbol{u}+1)0$ where $\boldsymbol{u} \in \mathcal{C}(000,010) $;
        the generating function for these words is  $y \cdot x^2\cdot \frac{1}{1-x-x^2}$,
\item[$-$] $\boldsymbol{w}  =0 101(\boldsymbol{u}+2)$ where $\boldsymbol{u} \in \mathcal{C}
      (000,010) $; the generating function for these words is  $y \cdot x^4\cdot \frac{1}{1-x-x^2}$.
\end{itemize}
Combining these cases and adding 1 corresponding to the empty word we have

$$
C_{\pi}(x,y)=1+x(1+x)\cdot \frac{1}{1-x-x^2}+yx^2(1+x^2)\left(\frac{1}{1-x-x^2}\right),
$$
which after calculation gives the desired result.
}
\endproof

\begin{Co}
\label{Co_000_021}
If $\pi=\{000,021\}$, then
$$
C_{\pi}(x)=- \frac{x^4+x^2+1}{x^2+x-1}\\
=1+x+3x^2+4x^3+8x^4+12x^5+20x^6+32x^7+O(x^6).
$$
\end{Co}

\begin{Pro}
\label{100_120}
If  $\pi=\{100,120\}$, then
$$
C_{\pi}(x,y)=-\frac{(x-1)^2}{x^3y-2x^2+3x-1}.
$$
\end{Pro}
\proof
\comm{
A word $\boldsymbol{w}\in\mathcal{C}(\pi)$ is in one of the following cases:
\begin{itemize}[parsep=-0.5mm,topsep=0mm]
\item[$-$] $\boldsymbol{w}$ is a w.i. Catalan word,
\item[$-$] $\boldsymbol{w}=\boldsymbol{u}(m-1)(\boldsymbol{v}+m)$ where $\boldsymbol{u}$ is a w.i. Catalan word other than $00\dotsb 0$, $m$ is the largest (last) entry of $\boldsymbol{u}$ and $\boldsymbol{v} \in C(\pi)$.
\end{itemize}
The generating function for the words of the first form is $\frac{1-x}{1-2x}$
and the generating function for the words of the second form is
$$
\left( \frac{1-x}{1-2x}-\frac{1}{1-x}\right)\cdot x\cdot y \cdot C_{\pi}(x,y).
$$
Combining these cases we deduce the functional equation below which solution gives the desired result:
$$C_{ \pi}(x,y)=\left(\frac{1-x}{1-2x}-\frac{1}{1-x}\right)\cdot x\cdot y \cdot C_{ \pi}(x,y)+  \frac{1-x}{1-2x}.
$$

}   
\endproof

\begin{Pro}
\label{110_120}
If $\pi=\{110,120\}$, then
$$
C_{\pi}(x,y)=-\frac{(x-1)^2}{x^3y-2x^2+3x-1}.
$$

\end{Pro}
\proof
\comm{
A word $\boldsymbol{w}\in\mathcal{C}(\pi)$ is in one of the following cases:
\begin{itemize}[parsep=-0.5mm,topsep=0mm]
\item[$-$] $\boldsymbol{w}$ is a w.i. Catalan word,
\item[$-$] $\boldsymbol{w}=\boldsymbol{u}(m+1)\boldsymbol{v}$ where $\boldsymbol{u}$ is a non-empty w.i. Catalan word, $m$ is the largest (last) entry of $\boldsymbol{u}$ and $\boldsymbol{v}$ is a word of the form 
$mm\cdots m(\boldsymbol{x}+m+1)$ with 
at least one $m$ in its prefix and $\boldsymbol{x} \in \mathcal{C}(\pi)$.
\end{itemize}
The generating function for the words of the first form is
$\frac{1-x}{1-2x}$.\\
For the second form, the generating function for the words $\boldsymbol{u}$
is $\frac{1-x}{1-2x}-1=\frac{x}{1-2x}$, and the generating function for the words
$mm\cdots m(\boldsymbol{x}+m+1)$ is $\frac{x}{1-x}\cdot C_{\pi}(x,y)$.
Thus, the generating function for the words of the second form is
$$
\frac{x}{1-2x}\cdot x\cdot y\cdot \frac{x}{1-x} \cdot \mathcal{C}_{\pi}(x,y).
$$
Combining these cases we deduce the functional equation
$$
C_{\pi}(x,y)=
\frac{x}{1-2x}\cdot x\cdot y \cdot \frac{x}{1-x}\cdot C_{\pi}(x,y)+\frac{1-x}{1-2x}.
$$

}
\endproof

The functional equations in the proofs of Propositions \ref{100_120} and
\ref{110_120} are different but the resulting bivariate generating functions are the same.
Instantiating $y$ by $1$ in $C_{\pi}(x,y)$ of these propositions
 we have the next corollary.
\begin{Co}
\label{Co_100_120}
If  $\pi=\{100,120\}$ or $\pi=\{110,120\}$, then
$$
C_{\pi}(x)=-\frac{(x-1)^2}{x^3-2x^2+3x-1}=
1+x+2x^2+5x^3+12x^4+28x^5+65x^6+151x^7+O(x^8),
$$

and $c_n(\pi)$ is the sequence $\href{https://oeis.org/A034943}{\tt A034943}$ in \cite{OEIS}.
\end{Co}

\begin{Pro}\label{021_110}
If $\pi=\{021,110\}$, then
$$
C_{\pi}(x,y)=-\frac{x^5y+x^4y-x^4-x^3y+4x^3-6x^2+4x-1}{(2x-1)(x-1)^3}.
$$
\end{Pro}
\proof
\comm{
A non-empty word $\boldsymbol{w}\in\mathcal{C}(\pi)$ is in one of the following cases:
\begin{itemize}[parsep=-0.5mm,topsep=0mm]
\item[$-$] $\boldsymbol{w}=0 \boldsymbol{u}$ where $\boldsymbol{u} \in \mathcal{C}(\pi)$; 
     the generating function for these words is $x \cdot C_{\pi}(x,y)$,
\item[$-$] $\boldsymbol{w}=0 (\boldsymbol{u}+1)$ where $\boldsymbol{u}$ is a non-empty w.i. Catalan word;
     the generating function for these words is  $x\cdot \frac{x}{1-2x}$,
\item[$-$] $\boldsymbol{w}=01\boldsymbol{u}$ where $\boldsymbol{u}$ is a non-empty w.i. Catalan word;
    the generating function for these words is  $x^2\cdot y \cdot \frac{x}{1-2x}$,
\item[$-$] $\boldsymbol{w}=\boldsymbol{u}0\dotsb 0$ where $\boldsymbol{u}$ is a s.i.  
     Catalan word of length at least three and $\boldsymbol{w}$ ending by at least one $0$;
     the generating function for these words is  $y \cdot \frac{x^4}{(1-x)^2}$.
\end{itemize}
Combining these cases and considering the empty word which contributes with $1$ to $C_{\pi}(x,y)$, we deduce the functional equation
$$
C_{\pi}(x,y)=1+x \cdot C_{\pi}(x,y)+x\cdot
\frac{x}{1-2x}+x^2\cdot y \cdot \frac{x}{1-2x}+y\cdot\frac{x^4}{(1-x)^2}.
$$

}
\endproof

\begin{Co}
\label{un_cor}
If $\pi=\{021,110\}$, then
$$
C_{\pi}(x)=
-\frac{x^5+3x^3-6x^2+4x-1}{(2x-1)(x-1)^3}
=1+x+2x^2+5x^3+12x^4+26x^5+53x^6+105x^7+O(x^8).
$$
\end{Co}

\begin{Pro}\label{110_201}
If $\pi=\{110,201\}$, then
\begin{equation*}
C_{\pi}(x,y)=  \frac{x^4y-x^3+3x^2-3x+1}{(x-1)(x^3y-2x^2+3x-1)}.
\end{equation*}

\end{Pro}
\proof
\comm{
A non-empty word $\boldsymbol{w}\in\mathcal{C}(\pi)$ is in one of the following cases:
\begin{itemize}[parsep=-0.5mm,topsep=0mm]
\item[$-$] $\boldsymbol{w}=0\boldsymbol{u}$ where $\boldsymbol{u} \in \mathcal{C}(\pi)$; 
      the generating function for these words is $x \cdot C_{\pi}(x,y)$,
\item[$-$] $\boldsymbol{w}=0(\boldsymbol{u}+1)$ where $\boldsymbol{u}$ is a non-empty word in 
      $\mathcal{C}(\pi)$; 
      the generating function for these words is $x \cdot (C_{\pi}(x,y)-1)$,
\item[$-$] $\boldsymbol{w}=\boldsymbol{u}0\dotsb 0$ where $\boldsymbol{u}$ is a s.i. 
      Catalan word of length at least $2$ and $\boldsymbol{w}$ ending by at least one $0$;  
      the generating function for these words is  $y \cdot\frac{x^3}{(1-x)^2}$,
\item[$-$] $\boldsymbol{w}=010\cdots 0(\boldsymbol{u}+1)$ where $\boldsymbol{u}$ is
      a non-empty word in $\mathcal{C}(\pi)$
      and $\boldsymbol{w}$ beginning by $010$;
      the generating function for these words is $y \cdot x^3 \cdot \frac{1}{1-x}\cdot (C_{\pi}(x)-1)$.
\end{itemize}
Combining these cases
and adding $1$ corresponding to the empty word we deduce the functional equation
$$   C_{\pi}(x,y)=1+x \cdot C_{\pi}(x,y)+x\cdot (C_{\pi}(x,y)-1)+
     y\cdot\frac{x^3}{(1-x)^2}+y\cdot x^3 \cdot \frac{1}{1-x}\cdot(C_{\pi}(x,y)-1).
$$
}
\endproof

\begin{Co}
\label{Cor_110_201}
If $\pi=\{110,201\}$, then
$$
C_\pi(x)=\frac{x^4-x^3+3x^2-3x+1}{(x-1)(x^3-2x^2+3x-1)}
=1+x+2x^2+5x^3+13x^4+32x^5+76x^6+178x^7+O(x^8).
$$
\end{Co}

\begin{Pro}\label{102_201} 
If $\pi=\{102,201\}$, then
\begin{equation*}
    \mathcal{C}_{\pi}(x,y)=  \frac{x^5y+x^4y^2-x^5-5x^4y+5x^4+6x^3y-10x^3-2x^2y+10x^2-5x+1}{(-x+1)(x^2y-x^2+2x-1)(x^2y-2x^2+3x-1)}.
\end{equation*}
\end{Pro}
\proof
\comm{
A non-empty word $\boldsymbol{w}\in\mathcal{C}(\pi)$ is in one of the following cases:
\begin{itemize}[parsep=-0.5mm,topsep=0mm]
\item[$-$] $\boldsymbol{w}=0 \boldsymbol{u}$ where $\boldsymbol{u}\in\mathcal{C}(\pi)$; the generating function for these words is  $x \cdot C_{\pi}(x,y)$,
\item[$-$] $\boldsymbol{w}=0 (\boldsymbol{u}+1)$ where $\boldsymbol{u}$ is a non-empty word in
           $\mathcal{C}(\pi) $; the generating function for these words is 
           $x \cdot (C_{\pi}(x,y)-1)$,
\item[$-$] $\boldsymbol{w}=0(\boldsymbol{v+1})0\dotsb 0$ where $\boldsymbol{u}$ is
           a non-empty word in $\mathcal{C}(\pi)$ and $\boldsymbol{w}$ ending by at least one $0$;  the generating function for these words is 
           $y \cdot x^2 \cdot  \frac{1}{1-x}\cdot(C_{\pi}(x,y)-1)$,

\item[$-$] $\boldsymbol{w}=01\cdots1\boldsymbol{u}$ where $\boldsymbol{u}$ is a binary word 
           beginning by a $0$ and different from $0\cdots 0$, or equivalently, 
           $\boldsymbol{u}$ a word in $\mathcal{C}(012)$ other than $0\cdots 0$; 
           the generating function for these words is $\frac{x^2}{1-x}\cdot y \cdot \left(C_{012}(x,y) - \frac{1}{1-x}\right)= \frac{x^2\cdot y}{1-x} \cdot \left(\frac{1-x+x^2-x^2y}{1-2x+x^2-x^2y} - \frac{1}{1-x}\right)$, see Theorem 4 in \cite{BKV}.
\end{itemize}
Combining these cases and adding $1$ corresponding to the empty word we deduce the functional equation
\begin{eqnarray*}
    C_{\pi}(x,y)=&&1+x \cdot C_{\pi}(x,y)+x\cdot (C_{\pi}(x,y)-1)+ \frac{x^2\cdot y}{1-x}\cdot
    (C_{\pi}(x,y)-1)\\
    &&+\frac{x^2\cdot y}{1-x} \cdot \left(\frac{1-x+x^2-x^2y}{1-2x+x^2-x^2y} - \frac{1}{1-x}\right).
\end{eqnarray*}

}
\endproof

\begin{Co}
\label{Co_102,201}
If $\pi=\{102,201\}$, then
$${C_{\pi}(x)} 
= \frac{x^4-4x^3+8x^2-5x+1}{(x-1)(2x-1)(x^2-3x+1)}
=1+x+2x^2+5x^3+14x^4+40x^5+113x^6+314x^7+O(x^8).
$$
\end{Co}
\begin{Pro}\label{100_110}
If $\pi=\{100,110\}$, then
$$
C_{\pi}(x,y)= \frac{x^4y-x^4+2x^3-2x+1}{(x-1)(x^3y-2x^3+x^2+2x-1)}.
$$
\end{Pro}
\proof
\comm{
A non-empty word in $\mathcal{C}(\pi)$ has one of the following forms:
\begin{itemize}[parsep=-0.5mm,topsep=0mm]
\item[$-$] $0 \boldsymbol{u}$ where $\boldsymbol{u} \in \mathcal{C}(\pi) $;
      the generating function for these words is  $x \cdot C_{\pi}(x,y)$,
\item[$-$] $0 (\boldsymbol{u}+1)$ where $\boldsymbol{u}$ is a non-empty
       word in $\mathcal{C}(\pi)$; 
      the generating function for these words is  $x \cdot (C_{\pi}(x,y)-1)$,
\item[$-$] $\boldsymbol{u}(m+1)(m+2)\boldsymbol{v}$ where $\boldsymbol{u}$ and 
      $\boldsymbol{v}$ are non-empty s.i. Catalan words, $m$ is the largest entry of $\boldsymbol{u}$
      and the length of $\boldsymbol{v}$ is less than or equal to that of $\boldsymbol{u}$;
      the generating function for these words is  $y \cdot x^4 \cdot(x+1) \cdot \frac{1}{(1-x^2)^2}$,
\item[$-$] $\boldsymbol{u}(m+1)\boldsymbol{u}(\boldsymbol{v}+m+1)$
      where $\boldsymbol{u}$ is a non-empty s.i. Catalan word, $m$ is the largest entry of 
      $\boldsymbol{u}$ and $\boldsymbol{v} \in \mathcal{C}(\pi) $;
      the generating function for these words is 
      $y \cdot x^3 \cdot \frac{1}{1-x^2}\cdot C_{\pi}(x,y)$.
\end{itemize}
Combining these cases and adding $1$ corresponding to the empty word we deduce the functional equation
$$
C_{\pi}(x,y)=1+x \cdot C_{\pi}(x,y)+x\cdot (C_{\pi}(x,y)-1)+  y \cdot x^4\cdot\frac{(x+1)}{(1-x^2)^2}+
 y\cdot x^3\cdot\frac{1}{1-x^2}\cdot C_{\pi}(x,y).
$$

}
\endproof

\begin{Co}
\label{Co_100_110}
If $\pi=\{100,110\}$, then
$$
C_{\pi}(x)=  \frac{-2x^3+2x-1}{(x-1)(x^3-x^2-2x+1)}=1+x+2x^2+5x^3+12x^4+28x^5+64x^6+145x^7+O(x^8).
$$
\end{Co}

\begin{Pro}\label{000_110}
If $\pi=\{000,110\}$, then
$$
C_{\pi}(x,y)=  \frac{x^3y+x^3-x^2-x+1}{(-x+1)(x^3-x^2y-x^2-x+1)}.
$$
\end{Pro}
\proof
\comm{
A non-empty word in $\mathcal{C}(\pi)$ has one of the following forms:
\begin{itemize}[parsep=-0.5mm,topsep=0mm]
\item[$-$] $0 (\boldsymbol{u}+1)$ where $\boldsymbol{u} \in \mathcal{C}(\pi) $;
     the generating function for these words is  $x \cdot C_{\pi}(x,y)$,
\item[$-$] $\boldsymbol{u}\boldsymbol{u}(\boldsymbol{v}+m+1)$ where 
     $\boldsymbol{u}$ is a non-empty s.i. Catalan word, $m$ is the largest (last) entry of 
     $\boldsymbol{u}$ and $\boldsymbol{v}\in \mathcal{C}(\pi)$; 
     the generating function for these words is  $y\cdot \frac{x^2}{1-x^2} \cdot 
     C_\pi(x,y)$, 
\item[$-$] $\boldsymbol{u}(m+1)\boldsymbol{v}$  where $\boldsymbol{u}$ 
      and $m$ are as above, and $\boldsymbol{v}$ is a non-empty s.i. Catalan word of length
      less than that of $\boldsymbol{u}$;
     the generating function for these words is $y \cdot x^3 \cdot (1+x)\cdot \frac{1}{(1-x^2)^2}$.
\end{itemize}
Combining these cases and adding $1$ corresponding to the empty word we deduce the functional equation
$$
C_\pi(x,y)=1+x \cdot C_{\pi}(x,y)+ y \cdot \frac{x^2}{1-x^2} \cdot C_{\pi}(x,y)+
y\cdot x^3 \cdot (1+x)\cdot \frac{1}{(1-x^2)^2}.
$$
}
\endproof

\begin{Co}
\label{Co_000_110}
If $\pi=\{000,110\}$, then
$$
C_{\pi}(x)=  \frac{2x^3-x^2-x+1}{(-x+1)(x^3-2x^2-x+1)}=
1+x+2x^2+4x^3+8x^4+15x^5+28x^6+51x^7+O(x^8).
$$
\end{Co}

\begin{Pro}\label{000_102}
If $\pi=\{000,102\}$ or $\pi=\{000,201\}$, then
$$
C_\pi(x,y)= \frac{yx^2-1}{yx^4+yx^2+x^2+x-1}.
$$
\end{Pro}
\proof
\comm{
If $\pi=\{000,102\}$, then 
a non-empty word in $\mathcal{C}(\pi)$ has one of the following forms:
\begin{itemize}[parsep=-0.5mm,topsep=0mm]
\item[$-$] $0(\boldsymbol{u}+1)$ where $\boldsymbol{u} \in \mathcal{C}(\pi) $; 
     the generating function for these words is  $x \cdot C_\pi(x,y)$,
\item[$-$] $00(\boldsymbol{u}+1)$ where $\boldsymbol{u} \in \mathcal{C}(\pi) $;
     the generating function for these words is  $x^2 \cdot C_\pi(x,y)$,
\item[$-$] $0(\boldsymbol{u}+1)0$ where $\boldsymbol{u}$ is a non-empty word in 
     $\mathcal{C}(\pi) $;
     the generating function for these words is $y \cdot x^2 \cdot (C_\pi(x,y)-1)$,
\item[$-$] $01(\boldsymbol{u}+2)01$ where $\boldsymbol{u} \in \mathcal{C}(\pi) $;
     the generating function for these words is  $y \cdot x^4 \cdot C_\pi(x,y)$.
\end{itemize}
Similarly, if $\pi=\{000,201\}$ and $\boldsymbol{w}$ is a non-empty word in $\mathcal{C}(\pi)$, then $\boldsymbol{w}$
has either one of the first three forms above, or 
\begin{itemize}[parsep=-0.5mm,topsep=0mm]
\item[$-$] $\boldsymbol{w}  =0 101(\boldsymbol{u}+2)$ where $\boldsymbol{u} \in \mathcal{C}(\pi) $;
      the generating function for these words is  $y \cdot x^4 \cdot C_{\pi}(x,y)$.
\end{itemize}
In both cases we obtain the functional equation
$$
C_\pi(x,y)=1+x\cdot C_\pi(x,y)+ x^2\cdot C_\pi(x,y)+y \cdot x^2\cdot (C_\pi(x,y)-1)+ y \cdot x^4\cdot C_\pi(x,y).
$$
}
\endproof
\begin{Co}
\label{Co_000_102}
If $\pi=\{000,102\}$ or $\pi=\{000,201\}$, then
$$
C_\pi(x)= \frac{x^2-1}{x^4+2x^2+x-1}=
1+x+2x^2+4x^3+9x^4+18x^5++38x^6+78x^7+O(x^8).
$$
\end{Co}

\begin{Pro}\label{000_120}
If $\pi=\{000,120\}$, then
$$
C_\pi(x,y)= - \frac{x^4y+x^3y+1}{x^4y+x^2+x-1}.
$$
\end{Pro}
\proof
\comm{
A non-empty word in $\mathcal{C}(\pi)$ has one of the following forms:
\begin{itemize}[parsep=-0.5mm,topsep=0mm]
\item[$-$] $0 (\boldsymbol{u}+1)$ where $\boldsymbol{u} \in \mathcal{C}(\pi) $;
     the generating function for these words is  $x \cdot C_\pi(x,y)$,
\item[$-$] $0 0(\boldsymbol{u}+1)$ where $\boldsymbol{u} \in \mathcal{C}(\pi) $;
     the generating function for these words is  $x^2 \cdot C_\pi(x,y)$,
\item[$-$] $0101 (\boldsymbol{u}+2)$ where $\boldsymbol{u} \in \mathcal{C}(\pi) $;
    the generating function for these words is  $y \cdot x^4 \cdot C_{\pi}(x,y)$.
\end{itemize}
Apart from these general cases, there are two other fixed length ones:
\begin{itemize}[parsep=-0.5mm,topsep=0mm]
\item[$-$] $0 10$; the corresponding generating function 
     is  $y \cdot x^3 $,
\item[$-$] $0 110$; the  corresponding generating function is  
     $ y \cdot x^4 $.
\end{itemize}
Combining these cases and adding $1$ corresponding to the empty word we deduce the functional equation
$$
C_\pi(x,y)=1+x\cdot C_\pi(x,y)+x^2\cdot C_{\pi}(x,y)+y \cdot x^4\cdot C_{\pi}(x,y)+ y \cdot x^3+y  \cdot x^4.
$$
}
\endproof

\begin{Co}
\label{Co_000_120}
If $\pi=\{000,120\}$, then
$$
C_\pi(x)= -\frac{x^4+x^3+1}{x^4+x^2+x-1}=1+x+2x^2+4x^3+8x^4+13x^5+23x^6+40x^7+O(x^8).
$$
\end{Co}

\begin{Pro}\label{201,210}
If $\pi=\{201,210\}$, then
$$
C_\pi(x,y)= \frac{x^4y-2x^4-3x^3y+7x^3+x^2y-9x^2+5x-1}{(2x-1)(x-1)(x^2y-2x^2+3x-1)}.
$$
\end{Pro}
\proof
\comm{
A non-empty word $\boldsymbol{w}$ in $\mathcal{C}(\pi)$ has one of the following forms:
\begin{itemize}[parsep=-0.5mm,topsep=0mm]
\item[$-$] $0 \boldsymbol{u}$ where 
     $\boldsymbol{u} \in \mathcal{C}(\pi) $; 
    the generating function for these words is $x \cdot C_\pi(x,y)$,
\item[$-$] $0 (\boldsymbol{u}+1$) where $\boldsymbol{u}$ is a non-empty
     word in $\mathcal{C}(\pi) $;
     the generating function for these words is  $x \cdot (C_\pi(x,y)-1)$,
\item[$-$] $0 1\cdots 1\boldsymbol{u}$ where 
     $\boldsymbol{u}$ is a non-empty word in  $\mathcal{C}(\pi)$ and $01$
     is a prefix of $\boldsymbol{w}$; the generating function for these words is
     $y \cdot \frac{x^2}{1-x} \cdot (C_\pi(x,y)-1)$,
\item[$-$] $0 (\boldsymbol{u}+1)0\dots 0 $ where $\boldsymbol{u}$ is a w.i.
     Catalan word other than $0\cdots 0$ and $\boldsymbol{w}$ ending by a $0$; 
     the generating function for these words is 
     $y\cdot \frac{x^2}{1-x} \cdot (\frac{1-x}{1-2x}-\frac{1}{1-x})$.
\end{itemize}
Combining these cases and adding 1 corresponding to the empty word we deduce the functional equation
$$C_\pi(x,y)=1+x\cdot C_\pi(x,y)+x\cdot (C_\pi(x,y)-1)+
y\cdot\frac{x^2}{1-x} \cdot (C_\pi(x,y)-1)+
y\cdot \frac{x^2}{1-x} \cdot \left( \frac{1-x}{1-2x}-\frac{1}{1-x} \right).
$$
}
\endproof

\begin{Co}
\label{Co_201,210}
If $\pi=\{201,210\}$, then
$$C_{\pi}(x)= \frac{x^4-4x^3+8x^2-5x+1}{(x-1)(2x-1)(x^2-3x+1)}=
1+x+2x^2+5x^3+14x^4+40x^5+113x^6+314x^7+O(x^8).
$$
\end{Co}

\begin{Pro}\label{102_210}
If $\pi=\{102,210\}$, then

$$
C_{\pi}(x,y)=
$$

$$
\frac{2x^7y-2x^7 +13x^6-10x^6y-x^6y^2-36x^5+19x^5y+55x^4-17x^4y-50x^3+7x^3y+27x^2-yx^2-8x+1}
{(x-1)^3(2x-1)^2(x^2y-x^2+2x-1)}.
$$
\end{Pro}
\proof
\comm{
A non-empty word $\boldsymbol{w}$ in $\mathcal{C}(\pi)$ has one of the following forms:
\begin{itemize}[parsep=-0.5mm,topsep=0mm]
\item[$-$] $0\boldsymbol{u}$ where $\boldsymbol{u}\in\mathcal{C}(\pi)$; 
   the generating function for these words is $x \cdot C_\pi(x,y)$,
\item[$-$] $0 (\boldsymbol{u}+1)$ where $\boldsymbol{u}$ is a non-empty word in 
   $\mathcal{C}(\pi)$;
   the generating function for these words is $x \cdot (C_\pi(x,y)-1)$,
\item[$-$] $01\cdots 1\boldsymbol{u}$ where $\boldsymbol{u}$ is a non-empty word in 
   $\mathcal{C}(012)$, 
  and $\boldsymbol{w}$ begins by $01$;
  the generating function for these words is 
  $y \cdot \frac{x^2}{1-x} \cdot (C_{012}(x,y)-1)=y \cdot \frac{x^2}{1-x} \cdot
  (\frac{1-x+x^2-x^2y}{1-2x+x^2-x^2y}-1)$ (see Theorem 4 in \cite{BKV} for the generating function
   of $C_{012}(x,y)$),
\item[$-$] $0 (\boldsymbol{u}+1)\boldsymbol{v} $ where $\boldsymbol{u}$ 
   is a w.i. Catalan word of length at least $2$ different from $0\cdots 0$ and $\boldsymbol{v}$
   is a non-empty word in $\mathcal{C}(010,012)$ (see Proposition \ref{seq_n}); 
   the generating function for these words is 
  $ x\cdot (\frac{1-x}{1-2x}-\frac{1}{1-x})\cdot  y \cdot C_{010,012}(x,y)  = y \cdot  x\cdot \frac{x}{(1-x)^2} \cdot (\frac{1-x}{1-2x}-\frac{1}{1-x})$.
\end{itemize}
Combining these cases and adding 1 corresponding to the empty word we deduce the functional equation
\begin{eqnarray*}
C_\pi(x,y)= &&1+x\cdot C_\pi(x,y)+x\cdot (C_\pi(x,y)-1)+y \cdot\frac{x^2}{1-x} \cdot \left(\frac{1-x+x^2-x^2y}{1-2x+x^2-x^2y}-1\right)+\\
& & y\cdot \frac{x^2}{(1-x)^2}\cdot\left(\frac{1-x}{1-2x}-\frac{1}{1-x}\right).
\end{eqnarray*}

}
\endproof

\begin{Co}
\label{Co_102,210}
If $\pi=\{102,210\}$, then
\begin{eqnarray*}
C_{\pi}(x)&=&\frac{2x^6 -17x^5+38x^4-43x^3+26x^2-8x+1}{(x-1)^3(2x-1)^3}\\
&=& 1+x+2x^2+5x^3+14x^4+40x^5+111x^6+295x^7+O(x^8).
\end{eqnarray*}
\end{Co}

\begin{Pro}\label{100,102}
If $\pi=\{100,102\}$, then

$$
C_\pi(x,y)=\frac{x^5y-x^4y-x^3+2x^3y+3x^2-x^2y-3x+1}{(x-1)(x^4y-x^3y-2x^2+x^2y+3x-1)}.
$$
\end{Pro}
\proof
\comm{
A non-empty word in $\mathcal{C}(\pi)$ has one of the following forms:
\begin{itemize}[parsep=-0.5mm,topsep=0mm]
\item[$-$] $0 \boldsymbol{u}$ where $\boldsymbol{u} \in \mathcal{C}(\pi) $;
     the generating function for these words is  $x\cdot C_\pi(x,y)$,
\item[$-$] $0 (\boldsymbol{u}+1)$ where $\boldsymbol{u}$
      is a non-empty word in $\mathcal{C}(\pi) $;
     the generating function for these words is  $x\cdot(C_\pi(x,y)-1)$,
\item[$-$] $0 (\boldsymbol{u}+1)0$ where $\boldsymbol{u}$ is as above;
     the generating function for these words is  $y\cdot x^2\cdot(C_\pi(x,y)-1)$,
\item[$-$] $011\cdots 1(\boldsymbol{u}+2)01$  where $\boldsymbol{u}$ is as above;
     the generating function for these words is
     $y\cdot\frac{x^4}{1-x}(C_\pi(x,y)-1)$,
\item[$-$] $\boldsymbol{u}\boldsymbol{v}$ where $\boldsymbol{u}$ and 
     $\boldsymbol{v}$ are binary words of length at least $2$ of the form $011\cdots 1$;
     the generating function for these words is $y\cdot\frac{x^4}{(1-x)^2}$.
\end{itemize}
Combining these cases and adding 1 corresponding to the empty word  we deduce the functional equation

$$
C_\pi(x,y)=1+x\cdot C_\pi(x,y)+x\cdot(C_\pi(x,y)-1)+y\cdot x^2\cdot(C_\pi(x,y)-1)+
y\cdot\frac{x^4}{1-x}(C_\pi(x,y)-1)+y\cdot\frac{x^4}{(1-x)^2}.
$$
}
\endproof

\begin{Co}
\label{Co_100_102}
If $\pi=\{100,102\}$, then
$$
C_\pi(x)=\frac{x^5-x^4+x^3+2x^2-3x+1}{(x-1)(x^4-x^3-x^2+3x-1)}
=1+x+2x^2+5x^3+13x^4+34x^5+87x^6+220x^7+O(x^8).
$$
\end{Co}

In the proof of the next proposition we need the following lemma where the generating functions
for some particular subsets of $\mathcal{C}(000,210)$ are given.

\begin{Le} 
\label{add_lem}
The bivariate generating function corresponding to 
\begin{enumerate}[parsep=-0.5mm,topsep=0mm]
\item the set $\mathcal{A}$ of words 
      $\boldsymbol{u}\boldsymbol{u}$ with $\boldsymbol{u}$ a non-empty s.i. Catalan word
      is $A(x,y)=x^2+yx^2\cdot\frac{x^2}{1-x^2}$ and $A(x)=\frac{x^2}{1-x^2}$;
\item the set $\mathcal{B}$ of words $\boldsymbol{u}\boldsymbol{v}$
      with $\boldsymbol{u}$ and $\boldsymbol{v}$ non-empty s.i. Catalan words and the length of 
      $\boldsymbol{v}$ is less than or equal to that of $\boldsymbol{u}$ is 
      $B(x,y)=y\cdot \frac{x^2}{1-x^2}\cdot \frac{1}{1-x}$;
\item the set $\mathcal{D}$ of words 
      $\boldsymbol{u}\boldsymbol{v}$ with $\boldsymbol{u}$ and 
      $\boldsymbol{v}$ non-empty s.i. Catalan words and the length of $\boldsymbol{v}$ is less
      than that of $\boldsymbol{u}$ is 
      $D(x,y)=y\cdot \frac{x^2}{1-x^2}\cdot \frac{x}{1-x}$.
\end{enumerate}
\end{Le} $ $
\proof
\comm{
1. For any even $n$ there is exactly one word of this form, so the monovariate 
   corresponding generating function is $\frac{x^2}{1-x^2}$; and only words of length 
   larger than two have one descent.

\noindent
2. The transformation
$(\boldsymbol{u}\boldsymbol{u},\boldsymbol{x})\mapsto \boldsymbol{u}\boldsymbol{x}\boldsymbol{u}$
where $\boldsymbol{u}\boldsymbol{u}\in\mathcal{A}$ and $\boldsymbol{x}$ is a s.i. 
Catalan word defines a bijection between pairs of such words and $\mathcal{B}$, and thus 
$B(x,y)=y\cdot A(x)\cdot\frac{1}{1-x}$.

\noindent
3. Similarly as point 2. 
}
\endproof

\begin{Pro}\label{000_210}
If $\pi=\{000,210\}$, then
$$
\mathcal{C}_{\pi}(x,y)= - \frac{(x+1)(x^3+x^3y-2x+1)}{(x^4-x^4y+x^3-2x^2-x+1)(x^2+x-1)}.
$$
\end{Pro}
\proof
\comm{
A non-empty word in $\mathcal{C}(000,210)$ has one of the following forms:
\begin{itemize}[parsep=-0.5mm,topsep=0mm]
\item[-] $\boldsymbol{u}$ with $\boldsymbol{u}\in\mathcal{D}$ and $\mathcal{D}$ as in Lemma \ref{add_lem}; 
         the generating function for these words is 
         $D(x,y)=y\cdot\frac{x^2}{1-x^2}\cdot \frac{x}{1-x}$,
\item[-] $\boldsymbol{u}(m+1)(m+1)(\boldsymbol{x}+m+2)\boldsymbol{v}$ with $\boldsymbol{u}$ and 
         $\boldsymbol{v}$ non-empty s.i. Catalan words and the length of 
         $\boldsymbol{v}$ is less than or equal to that of $\boldsymbol{u}$, $m$ is the largest entry of 
         $\boldsymbol{u}$, and $\boldsymbol{x}\in \mathcal{C}(000,010)$; the generating function for these
         words is 
         $B(x,y)\cdot x^2\cdot\frac{1}{1-x-x^2}=y\cdot\frac{x^2}{1-x^2}\cdot \frac{1}{1-x}\cdot x^2\cdot\frac{1}{1-x-x^2}$ 
          (see Lemma \ref{add_lem} and Proposition 
         \ref{Fibo}),
\item[-] $0(\boldsymbol{u}+1)$ where $\boldsymbol{u}\in\mathcal{C}(\pi)$; the generating function for these
         words is $x\cdot C_\pi(x,y)$,
\item[-] $\boldsymbol{u}\boldsymbol{u}(\boldsymbol{v}+m+1)$ where $\boldsymbol{u}$ is a non-empty 
         s.i. Catalan word,
         $m$ the largest entry of $\boldsymbol{u}$ and $\boldsymbol{v}\in\mathcal{C}(\pi)$; the generating
         function for these words is $A(x,y)\cdot C_\pi(x,y)$.
\end{itemize}
Combining these cases and adding $1$ corresponding to the empty 
word we deduce the functional equation
$$
C_\pi(x,y)=1+y\cdot\frac{x^2}{1-x^2}\cdot\frac{x}{1-x}+
y\cdot\frac{x^4}{1-x^2}\cdot \frac{1}{1-x}\cdot \frac{1}{1-x-x^2}+x\cdot C(x,y)+\left(x^2+y\cdot\frac{x^4}{1-x^2}\right)\cdot C(x,y).
$$
}
\endproof

\begin{Co}
\label{Co_000_210}
If $\pi=\{000,210\}$, then
$$
 C_\pi(x)=-  \frac{(x+1)(2x^3-2x+1)}{(x^3-2x^2-x+1)(x^2+x-1)}=
1+x+2x^2+4x^3+9x^4+18x^5+37x^6+72x^7+O(x^8)
.
$$
\end{Co}

In the proof of the next proposition we need the following lemma where the generating functions for 
two subsets of $\mathcal{C}(100,210)$ are given.

\begin{Le} $ $
\label{AB}
\begin{enumerate}[parsep=-0.5mm,topsep=0mm]
\item
The generating function corresponding to the set $\mathcal{\mathcal{E}}$ of words $\boldsymbol{u}\boldsymbol{v}$ with 
$\boldsymbol{u}$ a w.i. Catalan word, $\boldsymbol{v}$ a non-empty s.i. Catalan word and the largest 
entry of $\boldsymbol{v}$ is equal to that of $\boldsymbol{u}$ minus $1$ is 
$\frac{x^3}{(x-1)(x^2+x-1)}$.
\item
The generating function corresponding to the set $\mathcal{F}$ of words $\boldsymbol{u}\boldsymbol{v}$ 
with $\boldsymbol{u}$  a w.i. Catalan word, $\boldsymbol{v}$  a non-empty s.i. Catalan word and the largest 
entry of $\boldsymbol{v}$ is less than that of $\boldsymbol{u}$ minus $1$ is 
$\frac{x^3}{(x-1)(x^2+x-1)}\cdot\frac{x}{1-2x}$.
\end{enumerate}
\end{Le}
\proof
\comm{
1. If $\mathcal{E}_n$ is the set of words of length $n$ in $\mathcal{E}$, then $\mathcal{E}_n=\varnothing$ for $0\leq i\leq 2$,
$\mathcal{E}_3=\{010\}$ and $\mathcal{E}_4=\{0010,0110\}$. 
With $\boldsymbol{u}$ and $\boldsymbol{v}$ as above, for any $n\geq 3$, the transformation
\begin{itemize}[parsep=-0.5mm,topsep=0mm]
\item[] $\boldsymbol{u}\boldsymbol{v} \mapsto \boldsymbol{u}a\boldsymbol{v}$, with $a$ the maximal
   entry of $\boldsymbol{u}$, transforms a word in $\mathcal{E}_n$ into 
   one in $\mathcal{E}_{n+1}$ where the maximal entry occurs at least twice,
\item[] $\boldsymbol{u}\boldsymbol{v} \mapsto \boldsymbol{u}(a+1)\boldsymbol{v}(b+1)$, 
   with $a$ and $b$ the maximal entries of $\boldsymbol{u}$ and of $\boldsymbol{v}$ respectively, 
   transforms a word in $\mathcal{E}_n$ into 
   one in $\mathcal{E}_{n+2}$ where the maximal entry occurs once.
\end{itemize}

\noindent
Any word in $\mathcal{E}_n$, $n\geq 5$, except $0\cdots 010$, can be obtained in a unique way from either a word in $\mathcal{E}_{n-1}$ or in $\mathcal{E}_{n-2}$ by one of these transformations. This yields the recurrence
$|\mathcal{E}_n|=1+|\mathcal{E}_{n-1}|+|\mathcal{E}_{n-2}|$ for $n\geq 5$, and
the desired generating function is precisely that of the sequence $\left(|\mathcal{E}_n|\right)_{n\geq 0}$.

\noindent
2. Any pair of words $(\boldsymbol{w},\boldsymbol{x})$ with 
$\boldsymbol{w}=\boldsymbol{u}\boldsymbol{v}\in \mathcal{E}$ (with $\boldsymbol{u}$ and $\boldsymbol{v}$ as above) and $\boldsymbol{x}$
a non-empty w.i. Catalan word can be transformed into the word 
$\boldsymbol{u}\boldsymbol{x}\boldsymbol{v}\in \mathcal{F}$, and  
$(\boldsymbol{w},\boldsymbol{x})\mapsto \boldsymbol{u}\boldsymbol{x}\boldsymbol{v}$ is a bijection,
so the generating function for $\mathcal{F}$ is that for $\mathcal{E}$ multiplied by $\frac{x}{1-2x}$.

}
\endproof

\begin{Pro}\label{100,210}
\label{100_210}
If $\pi=\{100,210\}$, then
$$
C_\pi(x,y)=\frac{1-x}{1-2x}-\frac{x^3y}{(2x-1)(2x^3-x^3y+x^2-3x+1)}.
$$
\end{Pro}
\proof
\comm{
First we consider only words in $\mathcal{C}(\pi)$ having at least one descent, and we denote by 
$G(x,y)$ the corresponding generating function, and clearly $C_\pi(x,y)=\frac{1-x}{1-2x}+G(x,y)$.

A word in $\mathcal{C}(\pi)$ with at least one descent has one of the following forms:
\begin{itemize}[parsep=-0.5mm,topsep=0mm]
\item[$-$] $\boldsymbol{u}(\boldsymbol{\alpha}+s+1)(\boldsymbol{v}+s+t+1)$ where
$\boldsymbol{u}$ and $\boldsymbol{v}$ are both w.i. Catalan words, $\boldsymbol{\alpha}$ 
     belongs to the set $\mathcal{E}$ defined in Lemma \ref{AB}, and $s$ is the largest symbol of $\boldsymbol{u}$ (and for convenience $-1$ if $\boldsymbol{u}$ is empty) and $t$ that of $\boldsymbol{\alpha}$; the generating function for these words is  
     $y\cdot \frac{x^3}{(x-1)(x^2+x-1)}\cdot\left(\frac{1-x}{1-2x}\right)^2$,
\item[$-$] $\boldsymbol{u}(\boldsymbol{\alpha}+s+1)$ where
     $\boldsymbol{u}$ and $s$ are as above, and $\boldsymbol{\alpha}$ belongs to 
     the set $\mathcal{F}$ defined in Lemma \ref{AB}; the generating function for these words is  
     $y\cdot \frac{1-x}{1-2x}\cdot \frac{x^3}{(x-1)(x^2+x-1)}\cdot\frac{x}{1-2x}$,
\item[$-$] $\boldsymbol{u}(\boldsymbol{\alpha}+s+1)(\boldsymbol{v}+s+t+1)$ where
     $\boldsymbol{u}$ and $s$ are as above, $\boldsymbol{\alpha}$ belongs to $\mathcal{E}$, $\boldsymbol{v}$
     is a word in $\mathcal{C}(\pi)$ with at least one descent, and $t$ is the largest symbol of $\boldsymbol{\alpha}$; 
     the generating function for these words is  
     $y\cdot\frac{1-x}{1-2x} \cdot \frac{x^3}{(x-1)(x^2+x-1)}\cdot G(x,y)$.
\end{itemize}
It follows that $G(x,y)$ satisfies the functional equation
\begin{eqnarray*}
G(x,y)& = & y\cdot \frac{x^3}{(x-1)(x^2+x-1)}\cdot\left(\frac{1-x}{1-2x}\right)^2+
       y\cdot \frac{1-x}{1-2x}\cdot \frac{x^3}{(x-1)(x^2+x-1)}\cdot\frac{x}{1-2x}+\\
       &&y\cdot\frac{1-x}{1-2x} \cdot \frac{x^3}{(x-1)(x^2+x-1)}\cdot G(x,y).
\end{eqnarray*}
Finally, solving it and adding the generating function for the Catalan words with no descents
(that is, w.i. Catalan words) the statement holds.
}
\endproof
\begin{Co}
\label{Co_100_210}
If $\pi=\{100,210\}$, then
$$C_\pi(x)=\frac{x-1}{2x-1}-\frac{x^3}{(2x-1)(x-1)(x^2+2x-1)}=
1+x+2x^2+5x^3+13x^4+34x^5+88x^6+225x^7+O(x^8).
$$
\end{Co}

Numerical evidences let us believe that $c_n(100,210)$ is the sequence
\href{https://oeis.org/A267905}{\tt A267905} in \cite{OEIS}, however we failed to prove this formally.

\section{Final remarks}

Catalan words are in bijection with Dyck paths (see Figure \ref{Fig}) and thus
pattern avoiding Catalan words correspond to restricted Dyck paths.
For instance, a Catalan word avoiding $012$ corresponds to a Dyck path of height
at most two. In this context, it can be of interest to investigate
how our results on pattern avoiding Catalan words translate to 
corresponding restricted Dyck paths.

Even if in this article we restrict ourselves to the avoidance of two patterns of length 3,
some classes considered here can be trivially extended to larger length patterns, for instance $\mathcal{C}(102,201)=\mathcal{C}(01012,01201)$. 
In this light, it can be of interest to explore Catalan words avoiding patterns of length 4 or more,
triples of patterns or generalized patterns.

\begin{table}
{
\begin{center}
\begin{tabular}{|c|c|c|c|c|c|c|c|c|c|c|c|c|c|}
     \hline
$\sigma\backslash\tau$   & 000 & 001 & 010  & 011 & 012 & 021 & 100 & 101 & 102 & 110 & 120 & 201 & 210\\ \hline
000  & - & \cellcolor{blue!15} P. \ref{Fibo}&  P. \ref{Fibo} & $u.c$. & \cellcolor{blue!15}$u.c.$ & {\it C. \ref{Co_000_021}} & $s$ &
P. \ref{2_pw_n}& {\it C. \ref{Co_000_102}}& {\it C. \ref{Co_000_110}} & {\it C. \ref{Co_000_120}} & {\it C. \ref{Co_000_102}}& 
{\it C. \ref{Co_000_210}}\\ \hline 
001 & - & - & P. \ref{seq_n} & P. \ref{seq_n} & P. \ref{seq_n} & P. \ref{C2} & \cellcolor{blue!15} P. \ref{Fibom1}& $s$ & $s$ 
& P. \ref{C2} & P. \ref{C2} &\cellcolor{blue!15} $s$& P. \ref{C3}  \\ \hline 
010 & - & - & - & P. \ref{seq_n} & P. \ref{seq_n} & \cellcolor{blue!15} $s$ & $s$ & $s$ & $s$ & $s$ & $s$ & $s$ & $s$ \\ \hline 
011 & - & - & - & - & P. \ref{seq_n} & $s$ & P. \ref{2n1}&  $s$ &  $s$ &  $s$ & P. \ref{2n1} & $s$ & $s$\\ \hline 
012 & - & - & - & - & - & $s$ & P. \ref{C2} & P. \ref{C2}  & $s$ & P. \ref{C2}  & $s$ & $s$ & $s$ \\ \hline
021 & - & - & - & - & - & - &
P. \ref{2nmn} & 
P. \ref{2nmn} & 
 \cellcolor{blue!15} P. \ref{32n}  & {\it C. \ref{un_cor}} & P. \ref{np22n} & $s$ & $s$\\ \hline 
100 & - & - & - & - & - & - & - & P. \ref{Pell} & {\it C. \ref{Co_100_102}} & {\it C. \ref{Co_100_110}}&
C. \ref{Co_100_120}& P. \ref{Fibo_bis}& C. \ref{Co_100_210}\\ \hline 
101 & - & - & - & - & - & - & - & - & $s$ & 
P. \ref{2nmn} & 
P. \ref{2nmn} & $s$ & P. \ref{2nmp1}\\ \hline 
102 & - & - & - & - & - & - & - & - & - &
P. \ref{32n} &\cellcolor{blue!15} P. \ref{2nmp1}&  {\it C. \ref{Co_102,201}}& {\it C. \ref{Co_102,210}} \\ \hline
110 & - & - & - & - & - & - & - & - & - & - & C. \ref{Co_100_120} & {\it C. \ref{Cor_110_201}} & $s$\\ \hline 
120 & - & - & - & - & - & - & - & - & - & - & - & $s$ & $s$\\ \hline 
201 & - & - & - & - & - & - & - & - & - & - & - & - & {\it  C. \ref{Co_201,210}}\\ \hline 
210 & - & - & - & - & - & - & - & - & - & - & - & - & -\\ \hline 

\end{tabular}
\end{center}
}
\caption{\label{table2}
Pairs $\{\sigma,\tau\}$ where $\tau$ is superfluous for $\sigma$ are marked by $s$
and those yielding ultimately constant enumerating sequences (see Proposition \ref{UC}) by $u.c.$.
The references are to the propositions or the corollaries where the 
enumerating sequences or generating functions are given.
Pairs referred by the same proposition or corollary form a Wilf-equivalence class and
enumerating results that are not yet recorded in \cite{OEIS} are italicized.
Highlighted pairs are already enumerated in \cite{Baxter_Pudwell_15} in the context of ascent sequences, see Section \ref{sect_ascent}.}
\end{table}

\bigskip

\acknowledgements
\label{sec:ack}
The authors are grateful to the anonymous referees for their careful reading of the paper and their many insightful comments and suggestions.


\end{document}